\documentclass[11pt, a4paper]{article}
\pdfoutput=1

\usepackage[height=8.85in,width=6.75in]{geometry}

\usepackage{graphicx,rotating}     
\usepackage[bookmarksopen,colorlinks=true,linkcolor=light_blue,
citecolor=light_pink,urlcolor=dark_red,linktoc=all]{hyperref}

\usepackage{amsmath,tensor}
\usepackage{mathtools}
\usepackage{amsfonts}
\usepackage{amssymb}
\usepackage{slashed}
\usepackage{braket}
\usepackage{bm}
\usepackage{bbm}
\usepackage{cite}
\usepackage{comment}
\usepackage{mathrsfs}
\usepackage{xcolor}
\usepackage[utf8]{inputenc}
\usepackage[footnotesize]{caption}
\usepackage{bbold}
\usepackage{xfrac}
\usepackage{physics}
\usepackage{nicematrix}
\usepackage{multirow}

\usepackage{libertine}
\usepackage[ISO, upint]{libertinust1math}
\usepackage[scr=boondoxo,bb=boondox]{mathalpha}
\usepackage[T1]{fontenc}

\usepackage{chngcntr}
\counterwithin{equation}{section}

\usepackage{multicol}
\definecolor{dark_red}{rgb}{0.7, 0., 0.}
\definecolor{light_pink}{rgb}{1,0.4,0.4}
\definecolor{light_blue}{rgb}{0.284602,0.317763,0.963947}
\definecolor{cred}{RGB}{180,50,40}
\definecolor{darkgreen}{RGB}{0, 100, 0}
\definecolor{desy_blue}{HTML}{009EE2}
\definecolor{desy_orange}{HTML}{FD8800}
\definecolor{forestgreen}{HTML}{228B22}
\definecolor{ochre}{HTML}{CCAA2B}

\newcommand{\Mpl}{M_{\text{Pl}}}

\begin{document}

\hypersetup{pageanchor=false}
\begin{titlepage}

\begin{center}

\hfill KEK-TH-2733, RESCEU-15/25, IPMU25-0033, CTPU-PTC-25-25

\vskip .6in

{\Huge \bfseries
Torsion induced current-scalaron coupling\\[.1em]
in Einstein--Cartan gravity\\
}
\vskip .8in

{\LARGE Minxi He$^{a}$, Muzi Hong$^{b}$, Kyohei Mukaida$^{c}$}

\vskip .3in
\begin{tabular}{ll}
$^a$& \!\!\!\!\!\emph{Particle Theory and Cosmology Group, Center for Theoretical Physics of the Universe, }\\[-.3em]
& \!\!\!\!\!\emph{Institute for Basic Science (IBS),  Daejeon, 34126, Korea}\\
$^b$& \!\!\!\!\!\emph{Department of Physics, Graduate School of Science, }\\[-.3em]
& \!\!\!\!\!\emph{The University of Tokyo, Tokyo 113-0033, Japan}\\
$^b$& \!\!\!\!\!\emph{Research Center for the Early Universe (RESCEU), Graduate School of Science, }\\[-.3em]
& \!\!\!\!\!\emph{The University of Tokyo, Tokyo 113-0033, Japan}\\
$^b$& \!\!\!\!\!\emph{Kavli Institute for the Physics and Mathematics of the Universe (WPI), UTIAS,}\\[-.3em]
& \!\!\!\!\!\emph{The University of Tokyo, Kashiwa 277-8583, Japan}\\
$^c$& \!\!\!\!\!\emph{Theory Center, IPNS, KEK, 1-1 Oho, Tsukuba, Ibaraki 305-0801, Japan}\\
$^c$& \!\!\!\!\!\emph{Graduate University for Advanced Studies (Sokendai), }\\[-.3em]
& \!\!\!\!\!\emph{1-1 Oho, Tsukuba, Ibaraki 305-0801, Japan}
\end{tabular}

\end{center}
\vskip .6in

\begin{abstract}
\noindent
We investigate the matter current couplings with the scalar degrees of freedom originated from the torsion in Einstein--Cartan (EC) gravity. 
It has been shown in previous studies that the presence of the operators consisting of torsion components up to dimension four can naturally induce a (pseudo-)scalar degree of freedom, the scalaron. 
In this work, we consider the couplings between torsion and matter currents in this framework, and show that they can lead to couplings between these currents and the scalaron in the equivalent metric theory. 
We consider both gauge-invariant and gauge-dependent currents, showing general results and several concrete examples. 
These results are useful for the discussion of particle production processes after inflation in the EC framework, such as reheating and baryogenesis, and show the connection to the QCD $ \theta $ term. 
\end{abstract}

\end{titlepage}

\tableofcontents
\thispagestyle{empty}
\renewcommand{\thepage}{\arabic{page}}
\renewcommand{\thefootnote}{$\natural$\arabic{footnote}}
\setcounter{footnote}{0}
\newpage
\hypersetup{pageanchor=true}

\section{Introduction}

General relativity (GR) as the most successful classical theory of gravity has passed many experimental tests. The fundamental ingredient of GR is the metric tensor of the spacetime which solely determines the geometry of the spacetime, \textit{i.e.} the strength of gravity. 
The affine connection in GR is restricted to be the Levi-Civita connection uniquely determined by the metric. 
However, this is not a necessary condition for the construction of a gravity theory in a geometric way. 
Releasing this restriction leads to the general metric-affine gravity (see \textit{e.g.}, Ref.~\cite{Blagojevic:2012bc} and references therein) where the metric and affine connection are a priori independent, which allows additional geometric structures such as torsion and non-metricity in the affine connection. 
Recently, intensive attention has been drawn to a special type of metric-affine theory, the \textit{Einstein--Cartan (EC) gravity}, where the torsion is non-vanishing while non-metricity is kept zero. 

In the absence of matter coupling, an Eisntein--Hilbert action in EC formalism (and general metric-affine case) is equivalent to pure GR. 
The solution of the affine connection automatically turns out to be the Levi-Civita connection. 
Including torsion allows additional terms that are absent in GR, such as the Nieh--Yan (NY) term~\cite{Nieh:1981ww,Nieh:2008btw} which is topological, and the Holst term~\cite{Hojman:1980kv,Nelson:1980ph,Castellani:1991et,Holst:1995pc}.
These terms consist of torsion components.
If we add such operators up to dimension two, they do not affect the equation of motion because the solution of torsion is trivial. 
However, considering couplings between matter fields can lead to non-trivial results, although the torsion remains non-dynamical. 
In the context of inflation~\cite{Starobinsky:1980te,Sato:1980yn,Guth:1980zm,Mukhanov:1981xt,Linde:1981mu,Albrecht:1982wi},\footnote{See Ref.~\cite{Sato:2015dga} for a review.} for instance, the Higgs inflation (HI) in the EC framework~\cite{Shaposhnikov:2020gts,Shaposhnikov:2020frq} with a non-minimally coupled NY term provides a unified description of the metric~\cite{Cervantes-Cota:1995ehs,Bezrukov:2007ep,Barvinsky:2008ia} and the Palatini versions~\cite{Bauer:2010jg,Rasanen:2018ihz} of the Higgs inflation, which is realized by the modification of the kinetic structure of the Higgs fields from the solution of torsion. 
Given the unitarity issue in the metric HI~\cite{Ema:2016dny,Sfakianakis:2018lzf,Burgess:2009ea,Barbon:2009ya,Burgess:2010zq,Hertzberg:2010dc,Barvinsky:2009ii,Bezrukov:2010jz} and the Palatini HI~\cite{Bauer:2010jg,Ito:2021ssc}, 
EC HI allows a systematic study of the unitarity and ultraviolet (UV) extension of this class of models~\cite{He:2023vlj}.\footnote{See Refs.~\cite{Giudice:2010ka,Barbon:2015fla,Ema:2017rqn,Lee:2018esk,Koshelev:2020fok,Salvio:2015kka,Netto:2015cba,Liu:2018hno,Calmet:2016fsr,Ghilencea:2018rqg,Ema:2019fdd,Ema:2020zvg,Ema:2020evi,Gorbunov:2018llf,He:2018mgb} for the discussion of the UV extension of the metric HI. See also Refs.~\cite{Mikura:2021clt,He:2022xef,Mikura:2024fgp} for the discussion of UV extension of the Palatini HI and Ref.~\cite{Aoki:2022csb} in Weyl gravity. }
A systematic study of matter couplings in EC framework is carried out in Ref.~\cite{Karananas:2021zkl} where the authors consider operators up to dimension two in the gravity sector, dimension four in the matter sector, and dimension four in gravity-matter couplings. 
They coupled gauge-invariant currents to the torsion components.
It is found that the general torsion couplings with the currents can lead to interactions of such currents with themselves and other fields in the equivalent metric theory. 

On the other hand, even if there is no additional scalar field introduced, a (pseudo-)scalar degree of freedom, scalaron, can still be induced in a pure geometrical setup (see \textit{e.g.} \cite{Bombacigno:2019nua,Boudet:2023phd,Pradisi:2022nmh,DiMarco:2023ncs,He:2024wqv}, and \cite{Mikura:2023ruz} for a general discussion). 
For instance, by considering operators consisting of curvature and torsion up to dimension four~\cite{He:2024wqv},  
it is shown that inflation can be realized by means of the scalaron, including not only the well-known models such as the $ \alpha $-attractor~\cite{Ellis:2013nxa,Ferrara:2013rsa,Kallosh:2013yoa,Kallosh:2014rga,Carrasco:2015pla,Roest:2015qya,Linde:2015uga,Scalisi:2015qga},\footnote{See also pole inflation~\cite{Galante:2014ifa,Broy:2015qna,Terada:2016nqg,Fu:2022ypp,Aoki:2022bvj,Karamitsos:2021mtb,Pallis:2021lwk,Dias:2018pgj,Saikawa:2017wkg,Kobayashi:2017qhk} and running kinetic inflation~\cite{Takahashi:2010ky,Nakayama:2010kt}.} the Starobinsky model~\cite{Starobinsky:1980te}, and the quadratic chaotic inflation~\cite{Linde:1983gd}, but also other types such as deformation of the $ \alpha $-attractor model \cite{Salvio:2022suk, He:2024wqv}.  
One of the examples is the regularized pole inflation~\cite{He:2025bli} which can increase the predicted scalar spectral index $ n_s $ of curvature perturbation, reconciling the latest cosmic microwave background (CMB) observation results combining Atacama Cosmology Telescope (ACT)~\cite{ACT:2025fju,ACT:2025tim}, Planck~\cite{Planck:2018jri}, and BICEP/Keck~\cite{BICEP:2021xfz}.

The possible couplings between the scalaron and matter currents are also of interest because it can be essential for the discussion of particle production processes in the models in EC framework, such as (p)reheating~\cite{Starobinsky:1980te,Abbott:1982hn,Dolgov:1982th,Kofman:1994rk,Shtanov:1994ce,Kofman:1997yn,Greene:1997fu,Felder:2000hj,Felder:2001kt,Kofman:2001rb} and baryogenesis~\cite{sakharov:1967dj} in the inflation models mentioned above, etc. 
Reheating is an integrated part of inflationary cosmology~\cite{Linde:1981mu} which connects the inflation phase and the subsequent hot big bang and affects the $ e $-fold number of inflation~\cite{Liddle:2003as,Martin:2010kz} when contrasting the theoretical predictions with observation. 
Baryogenesis, on the other hand, is necessary to explain the observed baryon asymmetry~\cite{Planck:2018vyg}. 
Due to the presence of torsion in EC gravity, it has also been discussed that such couplings can be related to the strong CP problem in quantum chromodynamics (QCD)~\cite{Mielke:2006zp,Mercuri:2009zi,Mercuri:2009zt,Lattanzi:2009mg,Castillo-Felisola:2015ema,Karananas:2018nrj,Karananas:2024xja}, given that the scalaron arises as an axion from the torsion component and may gain an axion-like effective potential from coupling to the Chern--Simons term.

In this work, we investigate the couplings of matter currents with torsion in the EC framework with operators in the pure geometric part up to dimension four. 
As will be explained in the following, we treat the interested geometric quantities as independent pieces in the action, and consider only operators that form a complete square for the dimension-four part.  
It has been shown in \cite{He:2024wqv} that such combination gives a canonically normalizable (pseudo-)scalaron. 
After a brief introduction of EC gravity in Sec.~\ref{sec:pre}, we discuss the coupling of gauge-invariant currents in Sec.~\ref{sec:gauge-invariant} and list the asymptotic behavior in both large and small limits of scalaron field value.
Inflation possibilities out of the asymptotic region are also discussed.
We give two examples to illustrate the general results, and point out that only adding $\nabla_\mu \phi j_5^\mu$-type terms does not lead to solution of the strong CP problem (see \textit{e.g.} \cite{Kim:2008hd}).
In Sec.~\ref{sec:gauge-dependent}, we introduce the way of inclusion of gauge-dependent currents in our framework. 
We first show the general results and then discuss two examples. 
In the first, we show how the scalaron can be related to the $\theta$ term in QCD via coupling to the Chern--Simons term, while in the second we discuss the gauge field production during inflation.
In Sec.~\ref{sec:conclusion} we conclude and discuss future directions.

\section{Preliminaries of Einstein--Cartan gravity}
\label{sec:pre}

In this section, we briefly review the necessary basics of EC gravity which will be useful in the discussion throughout this paper. 

Since we are interested in the coupling between gravity and matter sector including fermions, it is useful to formulate the gravity sector in a way that local Lorentz symmetry can be seen explicitly. 
To achieve this, one can introduce vierbein fields $e^A_\mu(x)$ such that 
\begin{align}
    g_{\mu\nu} = e^A_\mu \, e^B_\nu \, \eta_{AB} ~,  
\end{align}
where the Greek indices denote general spacetime coordinates, while the Latin indices denote normal coordinates of the local inertia frame at spacetime point $x$. 
The former is raised and lowered by the spacetime metric $g_{\mu\nu}$, and the latter by the Minkowski metric $\eta_{AB}$.
Therefore, the quantities with capital Latin indices live in a locally flat spacetime where the Lorentz symmetry is manifest. 
To consistently write down the kinetic term of spinor fields in curved spacetime, we introduce the spin connection fields $\bar{\omega}^{AB}_\mu(x)$ such that the Dirac operator is given as 
\begin{equation}
    \bar{D}_\mu \equiv \partial_\mu - \frac{1}{8} \bar{\omega}^{AB}_\mu [\gamma_A,\gamma_B]~, 
\end{equation}
and $\bar{D}_A \equiv e_A^\mu \bar{D}_\mu$, with convention of $(-,+,+,+)$ and $\{\gamma_A,\gamma_B\}=-2 \eta_{AB}$. 
This allows $\bar{D}_A\psi$ to transform consistently under local Lorentz transformation using the normal coordinates at each spacetime point.

Curvature and torsion can also be expressed by vierbeins and spin connections as follows, respectively, 
\begin{equation}
    \bar{R}^{AB}{}_{\mu\nu} = \partial_\mu \bar{\omega}^{AB}_{\nu} -\partial_\nu \bar{\omega}^{AB}_{\mu} + \bar{\omega}^{A}_{\mu C} \bar{\omega}^{CB}_{\nu} - \bar{\omega}^{A}_{\nu C} \bar{\omega}^{CB}_{\mu}~,
\end{equation}
\begin{equation}
    T^A{}_{\mu\nu} = \partial_\mu e^A_{\nu} - \partial_\nu e^A_{\mu} + \bar{\omega}^A_{\mu B}e^B_\nu - \bar{\omega}^A_{\nu B}e^B_\mu~.
\end{equation}
The affine connection is related to vierbeins and spin connections as
$\bar{\Gamma}^\alpha_{\mu \nu} = e^\alpha_A (\partial_\mu e^A_\nu + \bar{\omega}^A_{\mu B}e^B_\nu)$, and Riemann curvature tensor and torsion tensor defined by affine connections are related to the definitions above as $\bar{R}^{\rho}{}_{\sigma\mu\nu}=e^{\rho}_A e^B_{\sigma}\bar{R}^A{}_{B \mu \nu}$ and $T^\lambda{}_{\mu\nu} = e^\lambda_AT^A{}_{\mu\nu}$.

In GR (or metric formalism), the metricity $\bar{\nabla}_\mu g_{\alpha \beta} = 0$ and the torsionless $ T^\lambda{}_{\mu\nu} =0 $ conditions are imposed. 
The former  requires the spin connection to satisfy $\bar{\omega}^{AB}_\mu = - \bar{\omega}^{BA}_\mu$, while the latter further restricts the spin connection to be determined by the vierbeins  
\begin{align}
    \omega^{AB}_{\mu} = \frac{1}{2} \qty[
    e^{\nu A}(\partial_\mu e^{B}_\nu - \partial_\nu e^{B}_\mu)
    - e^{\nu B}(\partial_\mu e^{A}_\nu - \partial_\nu e^{A}_\mu)
    - e_{\mu C} e^{\nu A} e^{\lambda B} (\partial_\nu e^{C}_\lambda - \partial_\lambda e^{C}_\nu)
    ]~,
\end{align}
where we use $\omega^{AB}_{\mu}$ to denote the spin connection in the metric formalism. 
The Levi-Civita connection is then expressed as $\Gamma^\alpha_{\mu \nu} = e^\alpha_A (\partial_\mu e^A_\nu + \omega^A_{\mu B}e^B_\nu)$. 
In EC gravity, on the other hand, torsionless condition is not imposed while metricity condition still holds, so the spin connections are not solely determined by the vierbeins.
Deviation of the connections from those in the metric formalism is expressed by the contorsion tensor:
\begin{equation}
    \bar{\omega}^{AB}_\mu = \omega^{AB}_\mu + K^{AB}_\mu, \quad
    \bar{\Gamma}^{\rho}_{\mu\nu} = \Gamma^{\rho}_{\mu\nu} + K^{\rho}{}_{\mu\nu}~,
\end{equation}
where $K^{AB}_\mu = e^{\alpha A} e^{\beta B} K_{\alpha \mu \beta}$. 
From the metricity condition, one finds $K^\lambda{}_{\mu\nu} = (1/2) (T^\lambda{}_{\mu\nu} + T_\nu{}^\lambda{}_\mu + T_\mu{}^\lambda{}_\nu)$. 
Using vector $T_\mu$, axial vector $S_\mu$, and tensor $q_{\alpha \beta \mu}$ components of the torsion tensor, the contorsion tensor is expressed as 
$K_{\alpha \mu \beta} = \frac{2}{3} T_{[\alpha} g_{\beta]\mu} - \frac{1}{12} E_{\alpha \mu \beta \nu} S^{\nu} - q_{\mu \beta \alpha}$.
See Appendix~\ref{app-EC} for details.

The Ricci scalar in EC gravity is connected to that in the metric formalism as follows:
\begin{equation}
     \bar{R} = R + 2 \nabla_\mu T^\mu - \frac{2}{3} T^\mu T_\mu + \frac{1}{24} S^\mu S_\mu + \frac{1}{2} q^{\mu \nu \rho} q_{\mu \nu \rho}.
     \label{eq-Rbar}
\end{equation}
With the fact in mind that the degrees of freedom of this theory can be divided to the vierbeins which contain all degrees of freedom of the metric formalism, and the contorsion $K^{AB}_\mu$ which are determined by vierbeins and torsions, we consider dimension-two operators $R$, $T^\mu T_\mu$, $S^\mu S_\mu$, $S^\mu T_\mu$, $\nabla_\mu T^\mu$ and $\nabla_\mu S^\mu$ as building blocks for the geometric part of our theory in the rest of our discussion.
Note that terms consist of two $q_{\mu \nu \rho}{}$'s always constrain themselves to zero in our setup, so we drop such terms from the beginning. 

The kinetic term of a spinor field minimally coupled to gravity in EC gravity is
\begin{equation}
    S_f = \int \sqrt{-g} \dd^4 x\,
    \qty( \frac{i}{2} \overline{\psi} \gamma^\mu \bar{D}_\mu \psi
    -  \frac{i}{2} \overline{ \bar{D}_\mu\psi} \gamma^\mu  \psi)~,
\end{equation}
where $\gamma^\mu \equiv e^{\mu}_A \gamma^A$.
This can be rewritten as
\begin{equation}
    S_f = \int \sqrt{-g} \dd^4 x\,
    \qty( \frac{i}{2} \overline{\psi} \gamma^\mu D_\mu \psi
    -  \frac{i}{2} \overline{ D_\mu\psi} \gamma^\mu  \psi + \frac{1}{8}j^{\mu}_5 S_\mu)~,
    \label{eq-naturalcoupling}
\end{equation}
by extracting the torsion part from the Dirac operator, 
with $j^{\mu}_5 \equiv \bar{\psi} \gamma^\mu \gamma_5 \psi$, $\gamma_5 \equiv i \gamma^0 \gamma^1 \gamma^2 \gamma^3 = -i \gamma_0 \gamma_1 \gamma_2 \gamma_3$, and
\begin{equation}
    D_\mu \equiv \partial_\mu - \frac{1}{8} \omega^{AB}_\mu [\gamma_A,\gamma_B]~.
\end{equation}
Thus, the coupling between the axial vector component of torsion and the chiral current of the fermion naturally arises with a coefficient $ 1/8 $. 

With these ingredients in mind, we can start to discuss more general couplings between torsion and matter in the following sections. 

\section{Gauge-invariant currents}
\label{sec:gauge-invariant}

In this section, we consider terms with gauge-invariant matter currents coupled to torsion components up to dimension four in the defining action. 
Our main interest is to study how the matter currents couple to the scalaron during and after single-field slow-roll inflation. 
Considering combinations of variables of $\bar{R}$, $\nabla_\mu S^\mu$, and $S_\mu T^\mu$ up to dimension four, the dimension-four part can be generally written in a quadratic form. 
As shown in Ref.~\cite{He:2024wqv}, if the symmetric matrix that specifies the quadratic form is of rank one, \textit{i.e.}, those operators form one complete square, the action is generally equivalent to that of a canonical scalar field, the scalaron, with an Einstein--Hilbert term. 
In the following, 
we consider the current couplings in the specific case where the dimension-four geometric part only consists of rank-one quadratic form. 
However, different from Ref.~\cite{He:2024wqv}, we take $ R (g) $, $S^\mu S_\mu$, $T^\mu T_\mu$, $\nabla_\mu S^\mu$, $\nabla_\mu T^\mu$ and $S_\mu T^\mu$ as independent variables of the quadratic form as discussed in the previous section.
For the general expression, we also freely choose the parameters in front of the linear geometric terms. 

In Sec.~\ref{sec:general1}, we start from a general action and derive the equivalent metric theories where the scalaron degree of freedom and its couplings with the gauge-invariant currents are explicitly shown. 
Then, in Sec.~\ref{sec:asymp} we discuss the asymptotic behaviors of the decay constants of these coupling terms as well as the current self-couplings, which indicate the behavior of the cutoff of the effective field theory (EFT).
We also discuss possible realizations of inflation out of the asymptotic region in Sec.~\ref{sec:noasymp}.
In Sec.~\ref{sec:example1}, we present two representative examples, the Starobinsky model and its deformation found in Ref.~\cite{He:2024wqv}, to illustrate the results in Sec.~\ref{sec:general1}. 
We also comment on the relation to the strong CP problem.

\subsection{General results of current couplings}
\label{sec:general1}

The general action satisfying the above conditions with gauge-invariant currents is as follows 
\begin{align}
    \label{eq-general1}
    \begin{split}
        S_\text{J} = \int \sqrt{-g_\text{J}} \dd^4 x\, & \left[
            \frac{\Mpl^2}{2} \qty(R_\text{J} + \beta_1 S_{\mu} S^{\mu} + \beta_2 T_{\mu} T^{\mu} + \beta_3 S_{\mu} T^{\mu} )
            \right.\\
            & 
            + \alpha_\text{R} \qty(\alpha R_\text{J} + \alpha_1 S_{\mu} S^{\mu} + \alpha_2 T_{\mu} T^{\mu} + \alpha_3 S_{\mu} T^{\mu} + \alpha_4 \nabla_\mu S^\mu + \alpha_5 \nabla_\mu T^\mu) ^ 2 \\
            & \left.
            + \sum_n \qty(\zeta_n S_\mu j_n^\mu + \xi_n T_\mu j_n^\mu)
            \right]~,
    \end{split}
\end{align}
where $j_n^\mu$'s are gauge-invariant matter currents. 
These currents are now directly coupled with different components of torsion in the third line. 
To have a scalaron degree of freedom, we generally require $\alpha_\text{R} \neq 0$ and at least one of the coefficients in the bracket in the second line is non-zero. 
Note that one of the parameters $\alpha_3$ and $\beta_3$ is redundant since we can always redefine $S^\mu$ or $T^\mu$ to shift one of them away.
For example, setting $S^{\prime \mu} = S^\mu + \beta_3 T^\mu / ( 2 \beta_1)$ can eliminate the term with $\beta_3$ in the case of $\beta_1 \neq 0$.
From now on we set $\alpha_3 = 0$ without loss of generality.
On the other hand, we retain the redundancy of $\alpha_R$ for the illustration of the examples in the next subsection.
Performing Legendre transformation, we obtain
\begin{align}
    \begin{split}
        S_\text{J} = \int \sqrt{-g_\text{J}} \dd^4 x\, & \Bigg[
            \frac{\Mpl^2}{2} \qty(R_\text{J} + \beta_1 S_{\mu} S^{\mu} + \beta_2 T_{\mu} T^{\mu} + \beta_3 S_{\mu} T^{\mu} )
            \\
            & 
            + \qty(\alpha R_\text{J} + \alpha_1 S_{\mu} S^{\mu} + \alpha_2 T_{\mu} T^{\mu} + \alpha_4 \nabla_\mu S^\mu + \alpha_5 \nabla_\mu T^\mu) 2 \chi -  \frac{\chi ^ 2}{\alpha_\text{R}}\\
            &
            + \sum_n \qty(\zeta_n S_\mu j_n^\mu + \xi_n T_\mu j_n^\mu)
            \Bigg]~,
    \end{split}
\end{align}
where $\chi$ is an auxiliary field.
By solving the constraint equation of $\chi$ and inserting the result back to the action, one can recover Eq.~\eqref{eq-general1}.
We further define $\Omega^2 = 1 + 4 \alpha \chi / \Mpl^2 > 0$ and solve the constraint equations for $S^\mu$ and $T^\mu$ respectively, to obtain the equivalent metric theory as 
\begin{align}
    \begin{split}
        S_\text{J} = \int \sqrt{-g_\text{J}} \dd^4 x\, & \Bigg\{
            \frac{\Mpl^2}{2} \Omega^2 R_\text{J} -\frac{\chi ^ 2}{\alpha_\text{R}}
            \\
            & 
            - \left[
             \qty(\frac{\Mpl ^ 2}{2} \beta_2 + 2 \alpha_2 \chi) \qty( \sum_n \zeta_n j_{n \mu} - 2 \alpha_4 \nabla_\mu \chi ) ^2 + \qty(\frac{\Mpl ^ 2}{2} \beta_1 + 2 \alpha_1 \chi) \qty( \sum_n \xi_n j_{n \mu} - 2 \alpha_5 \nabla_\mu \chi ) ^2 
            \right.\\
            & \quad \left.
            - \frac{\Mpl ^ 2}{2} \beta_3 \qty( \sum_n \zeta_n j_{n \mu} - 2 \alpha_4 \nabla_\mu \chi ) \qty(\sum_n \xi_n j_{n}^\mu - 2 \alpha_5 \nabla^\mu \chi) \right] \\
            & \times
            \qty[4 \qty(\frac{\Mpl ^ 2}{2} \beta_1 + 2 \alpha_1 \chi) \qty(\frac{\Mpl ^ 2}{2} \beta_2 + 2 \alpha_2 \chi) - \qty(\frac{\Mpl ^ 2}{2} \beta_3 ) ^ 2]^{-1}
            \Bigg\}~.
    \end{split}
    \label{eq-action1}
\end{align}

We do not further expand the above expression for now, but only summarize how to do Weyl transformation for each type of terms.
Let us consider the Weyl transformation from the Jordan frame to the Einstein frame:
\begin{equation}
    g_{\rm{E}\mu\nu} = \Omega^2 g_{\rm{J}\mu\nu}~.
\end{equation}
The first line of the above expression is transformed to
\begin{equation}
    \int \sqrt{-g_\text{E}} \dd^4 x\, 
    \qty[\frac{\Mpl^2}{2} R_\text{E} - \frac{3}{4} \Mpl^2 \frac{\nabla_\mu \Omega^2 \nabla^\mu \Omega^2}{\Omega^4} -V(\chi) ]~,~  V(\chi) \equiv \frac{\chi ^ 2}{\alpha_\text{R} \Omega ^ 4}  ~;
    \label{eq-kine1}
\end{equation}
terms proportional to $\nabla_\mu \chi \nabla^\mu \chi$ gain an extra $1 / \Omega^2$ factor:
\begin{equation}
    \int \sqrt{-g_\text{J}} \dd^4 x\,
    \nabla_\mu \chi \nabla^\mu \chi
    =
    \int \sqrt{-g_\text{E}} \dd^4 x\,
    \frac{\nabla_\mu \chi \nabla^\mu \chi}{\Omega^2}~;
\end{equation}
terms proportional to $\nabla_\mu \chi j^\mu$ do not obtain extra conformal factors:
\begin{equation}
    \int \sqrt{-g_\text{J}} \dd^4 x\,
    \nabla_\mu \chi j^\mu_{\text{J}}
    =
    \int \sqrt{-g_\text{E}} \dd^4 x\,
    \nabla_\mu \chi j^\mu_{\text{E}}~;
    \label{eq-conformal2}
\end{equation}
finally, terms proportional to $j_\mu j^\mu$ gain an extra $\Omega^2$ factor:
\begin{equation}
    \int \sqrt{-g_\text{J}} \dd^4 x\,
    j_{\text{J}\mu} j^\mu_{\text{J}}
    =
    \int \sqrt{-g_\text{E}} \dd^4 x\,
    \Omega^2 j_{\text{E}\mu} j^\mu_{\text{E}}~.
\end{equation}

For simplicity, we consider only one current, and analyze the asymptotic behavior of the ``decay constants" for possible phenomenological interest in future work. These results can be directly generalized to the case with more currents. 
The decay constants are defined as the inverse of the coefficient of the interaction terms that couple the derivative of the canonical scalaron and the matter current, and that of the self-coupling terms of the currents. 
Therefore, we replace $\sum_n \zeta_n j_{n \mu}$ and $\sum_n \xi_n j_{n \mu}$ with $\zeta j_{\mu}$ and $\xi j_{\mu}$, respectively, in the results above.
Then, Eq.~\eqref{eq-action1} with one matter current is written as 
\begin{align}
    \begin{split}
        S_\text{J} = \int \sqrt{-g_\text{J}} \dd^4 x\, & \Bigg\{
            \frac{\Mpl^2}{2} \Omega^2 R_\text{J} -\frac{\chi ^ 2}{\alpha_\text{R}}
            \\
            & 
            - \Bigg[
             \qty(\frac{\Mpl ^ 2}{2} \beta_2 + 2 \alpha_2 \chi) \qty( \zeta j_{\mu} - 2 \alpha_4 \nabla_\mu \chi ) ^2 + \qty(\frac{\Mpl ^ 2}{2} \beta_1 + 2 \alpha_1 \chi) \qty( \xi j_{\mu} - 2 \alpha_5 \nabla_\mu \chi ) ^2
            \\
            & \quad
            - \frac{\Mpl ^ 2}{2} \beta_3 \qty( \zeta j_{\mu} - 2 \alpha_4 \nabla_\mu \chi ) \qty(\xi j^{\mu} - 2 \alpha_5 \nabla^\mu \chi) \Bigg] \\
            &\times
            \qty[4 \qty(\frac{\Mpl ^ 2}{2} \beta_1 + 2 \alpha_1 \chi) \qty(\frac{\Mpl ^ 2}{2} \beta_2 + 2 \alpha_2 \chi) - \qty(\frac{\Mpl ^ 2}{2} \beta_3 ) ^ 2]^{-1}
            \Bigg\}~.
    \end{split}
\end{align}
We focus on the terms proportional to $\nabla_\mu \chi \nabla^\mu \chi$,  $\nabla_\mu \chi j^\mu$, and $j^\mu j_\mu$ in the last three lines. 
These terms can be written as 
\begin{align}
    \mathcal{L}_{\text{J}} \supset - F(\chi) \frac{\nabla_\mu \chi \nabla^\mu \chi}{2} - G(\chi) \nabla_\mu \chi j^\mu - I(\chi) j^\mu j_\mu~, 
\end{align}
where 
\begin{align}
    F(\chi) &= \frac{8 ( \alpha_2 \alpha_4^2 + \alpha_1 \alpha_5^2 ) \chi + 2\Mpl^2 ( \beta_2 \alpha_4^2  - \beta_3 \alpha_4 \alpha_5 + \beta_1 \alpha_5^2 )}
    {8 \alpha_1 \alpha_2 \chi^2 + 2 \Mpl^2 ( \alpha_1 \beta_2 + \alpha_2 \beta_1) \chi + \frac{\Mpl^4}{8} (4 \beta_1 \beta_2 - \beta_3^2)}~, \nonumber \\
    G(\chi) &= \frac{- 4(\alpha_2 \alpha_4 \zeta + \alpha_1 \alpha_5\xi) \chi + \frac{\Mpl^2}{2} (- 2 \beta_2 \alpha_4 \zeta + \beta_3 \alpha_5 \zeta + \beta_3 \alpha_4 \xi - 2 \beta_1 \alpha_5 \xi )}
    {8 \alpha_1 \alpha_2 \chi^2 + 2 \Mpl^2 (\alpha_1 \beta_2 + \alpha_2 \beta_1 ) \chi + \frac{\Mpl^4}{8} (4 \beta_1 \beta_2 - \beta_3^2)} ~, \label{eq-Fchi} \\
    I(\chi) &= \frac{\qty(\alpha_1 \xi^2 + \alpha_2 \zeta^2) \chi + \frac{\Mpl^2}{4} \qty(\beta_1 \xi^2 - \beta_3 \zeta \xi + \beta_2 \zeta^2)}{8 \alpha_1 \alpha_2 \chi^2 + 2 \Mpl^2 (\alpha_1 \beta_2 + \alpha_2 \beta_1 ) \chi + \frac{\Mpl^4}{8} (4 \beta_1 \beta_2 - \beta_3^2)} \nonumber~.
\end{align}
After Weyl transformation, these terms become
\begin{align}
    \mathcal{L}_{\text{E}} \supset - \frac{F(\chi)}{\Omega^2} \frac{\nabla_\mu \chi \nabla^\mu \chi}{2} - G(\chi) \nabla_\mu \chi j^\mu - \Omega^2 I(\chi) j^\mu j_\mu~.
    \label{eq-kine2}
\end{align}
Here, the currents should be understood as the rescaled ones in the Einstein frame, but we omit the subscript ``E'' for convenience. 

We combine Eqs.~\eqref{eq-kine1} and \eqref{eq-kine2} to write down the kinetic term of $\chi$ in the Einstein frame 
\begin{equation}
    \mathcal{L}_{\text{kinetic}} = - \qty(\frac{24 \alpha^2}{\Omega^4 \Mpl^2} + \frac{F(\chi)}{\Omega^2}) \frac{\nabla_\mu \chi \nabla^\mu \chi}{2}~.
    \label{eq-kine}
\end{equation}
Here, we implicitly assume the coefficients are properly chosen such that, in the certain field range of interest, the terms in the bracket as a whole is positive-definite and the kinetic term has a proper sign. 
Note that one can consider the contribution of $F(\chi)$ to the kinetic term independently from the contribution of the Weyl transformation.
If $\alpha = 0$, \textit{i.e.}, $\Omega^2 = 1$, there is no need to do Weyl transformation, whereas one still has the contribution of $F(\chi)$ to the kinetic term.
We define the canonicalized inflaton field $ \phi $ as
\begin{equation}
    \phi \equiv \int \sqrt{\frac{24 \alpha^2}{\Omega^4 \Mpl^2} + \frac{F(\chi)}{\Omega^2}} \dd \chi~,
    \label{eq-cano}
\end{equation}
and formally write down $\chi$ as a function of $\phi$ in the form $\chi(\phi)$.  
After canonicalization, the inflaton-current coupling and current self-coupling can then be expressed as 
\begin{equation}
    \mathcal{L}_{\text{E}} \supset - \frac{ G(\chi(\phi))}{\sqrt {\frac{24 \alpha^2}{\Omega^4 \Mpl^2} + \frac{F(\chi(\phi))}{\Omega^2}}} \nabla_\mu \phi j ^ \mu - \Omega^2 I(\phi(\chi)) j^\mu j_\mu ~, 
    \label{eq-coupling_gi}
\end{equation}
from which the decay constants are then read as 
\begin{equation}
    f_1(\phi) = \frac{1}{G(\chi(\phi))} \sqrt {\frac{24 \alpha^2}{\Omega^4 \Mpl^2} + \frac{F(\chi(\phi))}{\Omega^2}} ~, \quad f_2(\phi) = \frac{1}{\Omega^2 I(\chi)} ~.
\end{equation}

\subsubsection{Asymptotic behavior}
\label{sec:asymp}

Now we focus on the cases that for positive $\chi$, the quantity in bracket of Eq.~\eqref{eq-kine} is positive and neither numerator nor denominator has zero-points.
We analyze the behavior of the decay constant in asymptotic regions, \textit{i.e.}, $\chi \to \infty$\footnote{If instead considering $\chi < 0$ and $\chi \to -\infty$, one may flip the sign of $\alpha_1$ and $\alpha_2$ to obtain the same result.} and $\chi \to 0$.
Let us first consider $\chi \to \infty$.
From Eqs.~\eqref{eq-Fchi} and \eqref{eq-kine}, one can see the coefficient of $\nabla_\mu \chi \nabla^\mu \chi / 2$ typically scales as $\sim 1 / \chi^2$, assuming $\alpha_2 \alpha_4^2 + \alpha_1 \alpha_5^2 \neq 0$, $\alpha_1 \alpha_2 \neq 0$ and $\alpha \neq 0$.
$G(\chi)$ scales as $\sim 1 / \chi$ if $\alpha_2 \alpha_4 \zeta + \alpha_1 \alpha_5\xi \neq 0$ and $I(\chi)$ scales as $\sim 1 / \chi$ if $ \alpha_1 \xi^2 +  \alpha_2 \zeta^2 \neq 0$, and thus $f_1$ and $f_2$ are constant. 

For specific combinations of parameters, $F(\chi)$, $G(\chi)$ and $I(\chi)$ may show other behaviors when taking $\chi \to \infty$.
The results are summarized in Table~\ref{table-inf} for $\alpha=0$ and Table~\ref{table-infa} for $\alpha \neq 0$.
The asymptotic behavior of the potential $ V(\chi (\phi)) $ is also listed.
Note that if $G(\chi)=0$ or $I(\chi)=0$, the coupling between inflaton and the current or the self-coupling of currents vanish, respectively, and thus there is no corresponding decay constants.
Most of the cases with vanishing couplings are not shown in the tables since the results are trivial.

Similarly, in the asymptotic region of $\chi \to 0$, the coefficient of $\nabla_\mu \chi \nabla^\mu \chi / 2$ is constant if $\beta_2 \alpha_4^2  - \beta_3 \alpha_4 \alpha_5 + \beta_1 \alpha_5^2 \neq 0$ and $4 \beta_1 \beta_2 - \beta_3^2 \neq 0$.
$G(\chi)$ and $I(\chi)$ and therefore $f_1$ and $f_2$ are also constant assuming $- 2 \beta_2 \alpha_4 \zeta + \beta_3 \alpha_5 \zeta + \beta_3 \alpha_4 \xi - 2 \beta_1 \alpha_5 \xi \neq 0$ and $ \beta_1 \xi^2 - \beta_3 \zeta \xi +  \beta_2 \zeta^2 \neq 0$.
Other possibilities are summarized in Table~\ref{table-0} for $\alpha = 0$ and Table~\ref{table-0a} for $\alpha \neq 0$.

Note that not all the cases listed in the tables are well-defined field theories.
The decay constants indicate the cutoff of EFT.
Thus, cases where the decay constant goes to 0, \textit{i.e.}, $f \sim \phi^{n}$ for $\phi \to 0$ should actually be excluded when discussing reheating process.
One also needs to make sure the inflation energy scale is below the cutoff of EFT when discussing inflation.

\NiceMatrixOptions{cell-space-limits=4pt}
\begin{table}
    \caption{Behavior of $f_1$ and $f_2$ at $\chi \to \infty$ for $\alpha = 0$. In this and the following tables $\Mpl$ is taken to be 1. The asymptotic behavior of the potential is also listed.}
\begin{center}
\begin{NiceTabular}{|c|c|c|c|}\hline
    $F(\chi) \sim,\,G(\chi) \sim,\, I(\chi) \sim$ & $f_1(\phi) \sim$ & $f_2(\phi) \sim$ & $V(\phi) \sim$ \\ \hline

    $\chi,\,\chi,\,\chi$ & $\phi^{-\frac{1}{3}}$ & $\phi^{-\frac{2}{3}}$ & $\phi^\frac{4}{3}$ \\ \hline
    $\chi,\,$const.,$\,0$ & $\phi^{\frac{1}{3}}$ & None & $\phi^\frac{4}{3}$ \\ \hline

    const.,$\,$const.,$\,$const. & const. & const. & $\phi^2$ \\ \hline
    const.,$\,\chi^{-1},\,\chi^{-1}$ & $\phi$ & $\phi$ & $\phi^2$ \\ \hline

    $\chi^{-1},\,\chi^{-1},\,$const. & $\phi$ & const. & $\phi^4$ \\ \hline
    $\chi^{-1},\,\chi^{-1},\,\chi^{-1}$ & $\phi$ & $\phi^2$ & $\phi^4$ \\ \hline
    $\chi^{-1},\,\chi^{-1},\,\chi^{-2}$ & $\phi$ & $\phi^4$ & $\phi^4$ \\ \hline
    $\chi^{-1},\,\chi^{-2},\,\chi^{-1}$ & $\phi^3$ & $\phi^2$ & $\phi^4$ \\ \hline

    $\chi^{-2},\,\chi^{-1},\,\chi^{-1}$ & const. & $ e^{\phi}$ & $ e^{\phi}$  \\ \hline
    $\chi^{-2},\,\chi^{-1},\,\chi^{-2}$ & const. & $ e^{\phi}$ & $ e^{\phi}$ \\ \hline
    $\chi^{-2},\,\chi^{-2},\,\chi^{-2}$ & $ e^{\phi}$ & $ e^{\phi}$ & $ e^{\phi}$  \\ \hline

\end{NiceTabular}
\end{center}
\label{table-inf}
\end{table}
\NiceMatrixOptions{cell-space-limits=4pt}
\begin{table}
    \caption{Behavior of $f_1$ and $f_2$ at $\chi \to \infty$ for $\alpha \neq 0$. The potential behaves as $V(\phi) \sim \Mpl^4 \cdot \text{const.}$}
\begin{center}
\begin{NiceTabular}{|c|c|c|}\hline
    $F(\chi) \sim,\,G(\chi) \sim,\, I(\chi) \sim$ & $f_1(\phi) \sim$ & $f_2(\phi) \sim$ \\ \hline

    $\chi,\,\chi,\,\chi$ & $\phi^{-1}$ & $\phi^{-2}$ \\ \hline
    $\chi,\,$const.,$\,0$ & const. & None  \\ \hline

    const.,$\,$const.,$\,$const. & $\phi^{-1}$ & $\phi^{-2}$ \\ \hline
    const.,$\,\chi^{-1},\,\chi^{-1}$ & $\phi$ & const. \\ \hline

    $\chi^{-1},\,\chi^{-1},\,$const. & const. & $ e^{-\phi}$ \\ \hline
    $\chi^{-1},\,\chi^{-1},\,\chi^{-1}$ & const. & const. \\ \hline
    $\chi^{-1},\,\chi^{-1},\,\chi^{-2}$ & const. & $ e^{\phi}$ \\ \hline
    $\chi^{-1},\,\chi^{-2},\,\chi^{-1}$ & $ e^{\phi}$ & const. \\ \hline

    $\chi^{-2},\,\chi^{-1},\,\chi^{-1}$ & const. & const. \\ \hline
    $\chi^{-2},\,\chi^{-1},\,\chi^{-2}$ & const. & $ e^{\phi}$ \\ \hline
    $\chi^{-2},\,\chi^{-2},\,\chi^{-2}$ & $ e^{\phi}$ & $ e^{\phi}$ \\ \hline

\end{NiceTabular}
\end{center}
\label{table-infa}
\end{table}
\NiceMatrixOptions{cell-space-limits=4pt}
\begin{table}
    \caption{Behavior of $f_1$ and $f_2$ at $\chi \to 0$ for $\alpha = 0$. The asymptotic behavior of the potential is also listed.}
\begin{center}
\begin{NiceTabular}{|c|c|c|c|}\hline
    $F(\chi) \sim,\,G(\chi) \sim,\, I(\chi) \sim$ & $f_1(\phi) \sim$ & $f_2(\phi) \sim$ & $V(\phi)\sim$  \\ \hline

    $\chi,\,\chi,\,\chi$ & $\phi^{-\frac{1}{3}}$ & $\phi^{-\frac{2}{3}}$ & $\phi^{\frac{4}{3}}$ \\ \hline
    $\chi,\,$const.,$\,\chi$ & $\phi^{\frac{1}{3}}$ & $\phi^{-\frac{2}{3}}$ & $\phi^{\frac{4}{3}}$ \\ \hline
    $\chi,\,$const.,$\,$ const. & $\phi^{\frac{1}{3}}$ & const. & $\phi^{\frac{4}{3}}$ \\ \hline

    const.,$\,\chi,\,$const. & $\phi^{-1}$ & const. & $\phi^2$ \\ \hline
    const.,$\,$const.,$\,\chi$ & const. & $\phi^{-1}$ & $\phi^2$ \\ \hline
    const.,$\,$const.,$\,$const. & const. & const. & $\phi^2$ \\ \hline
    const.,$\,$const.,$\,\chi^{-1}$ & const. & $\phi$ & $\phi^2$ \\ \hline

    $\chi^{-1},\,$const.$\,$const. & $\phi^{-1}$ & const. & $\phi^4$ \\ \hline
    $\chi^{-1},\,\chi^{-1},\,\chi^{-1}$ & $\phi$ & $\phi^2$ & $\phi^4$ \\ \hline

    $\chi^{-2},\,\chi^{-1},\,0$ & const. & None & $ e^{\phi}$  \\ \hline
    $\chi^{-2},\,\chi^{-2},\,\chi^{-2}$ & $ e^{\phi}$ & $ e^{\phi}$  & $ e^{\phi}$  \\ \hline

\end{NiceTabular}
\end{center}
\label{table-0}
\end{table}
\NiceMatrixOptions{cell-space-limits=4pt}
\begin{table}
    \caption{Behavior of $f_1$ and $f_2$ at $\chi \to 0$ for $\alpha \neq 0$. The asymptotic behavior of the potential is also listed.}
\begin{center}
\begin{NiceTabular}{|c|c|c|c|}\hline
    $F(\chi) \sim,\,G(\chi) \sim,\, I(\chi) \sim$ & $f_1(\phi) \sim$ & $f_2(\phi) \sim$ & $V(\phi)\sim$  \\ \hline

    $\chi,\,\chi,\,\chi$ & $\phi^{-1}$ & $\phi^{-1}$ & $\phi^{2}$ \\ \hline
    $\chi,\,$const.,$\,\chi$ & const. & $\phi^{-1}$ & $\phi^{2}$ \\ \hline
    $\chi,\,$const.,$\,$ const. & const. & const. & $\phi^{2}$ \\ \hline

    const.,$\,\chi,\,$const. & $\phi^{-1}$ & const. & $\phi^2$ \\ \hline
    const.,$\,$const.,$\,\chi$ & const. & $\phi^{-1}$ & $\phi^2$ \\ \hline
    const.,$\,$const.,$\,$const. & const. & const. & $\phi^2$ \\ \hline
    const.,$\,$const.,$\,\chi^{-1}$ & const. & $\phi$ & $\phi^2$ \\ \hline

    $\chi^{-1},\,$const.$\,$const. & $\phi^{-1}$ & const. & $\phi^4$ \\ \hline
    $\chi^{-1},\,\chi^{-1},\,\chi^{-1}$ & $\phi$ & $\phi^2$ & $\phi^4$ \\ \hline

    $\chi^{-2},\,\chi^{-1},\,0$ & const. & None & $ e^{\phi}$  \\ \hline
    $\chi^{-2},\,\chi^{-2},\,\chi^{-2}$ & $ e^{\phi}$ & $ e^{\phi}$  & $ e^{\phi}$ \\ \hline

\end{NiceTabular}
\end{center}
\label{table-0a}
\end{table}

\subsubsection{Discussion of inflation possibilities}
\label{sec:noasymp}

In the asymptotic region of $\chi \to \infty$, it is possible for $\phi$ to serve as inflaton and realize inflation.
One can notice from the last column in Table~\ref{table-inf} that it is possible to realize inflation with power-law potential in the case of $\alpha = 0$.
And in the case of $\alpha \neq 0$, asymptotically we have a flat potential, which is also possible to realize inflation.

Apart from the behavior in the asymptotic region, there are other possibilities to realize inflation due to the property of the  kinetic term Eq.~\eqref{eq-kine}.
In this section we briefly discuss several cases of inflation realized not in the asymptotic region.

First note that the denominator and the numerator of the term in the bracket of Eq.~\eqref{eq-kine} is a polynomial of $\chi$.
In the previous section, we considered cases that the quantity in the bracket is positive for $0 < \chi < \infty$.
Generally, the numerator or the denominator can have zero points.
The $\chi$ field is in the range such that the quantity in the bracket is positive, so that the kinetic term has a positive sign.
If the denominator of the bracket in Eq.~\eqref{eq-kine} has zero point(s), the potential becomes flat near it, which might be suitable for inflation\footnote{See \cite{Galante:2014ifa,Broy:2015qna,Terada:2016nqg,Fu:2022ypp,Aoki:2022bvj,Karamitsos:2021mtb,Pallis:2021lwk,Dias:2018pgj,Saikawa:2017wkg,Kobayashi:2017qhk} for pole inflation.
}.
To illustrate the case when the denominator has zero point(s), we consider an example with $\alpha = 0$ and $\alpha_2 \alpha_4^2 + \alpha_1 \alpha_5^2 = 0$, and without loss of generality we consider the kinetic term and potential of $\chi$ as follows:
\begin{equation}
    \mathcal{L}_{\chi} = - \frac{a}{\chi(\chi-b)} \frac{\nabla_\mu \chi \nabla^\mu \chi}{2} - \frac{(\chi - c)^2}{\alpha_{\rm{R}}}~,
\end{equation}
where $ b>0 $. 
Now let us first consider the case of $a>0$ and $\chi >b$.
To canonically normalize $\chi$, we perform the field redefinition:
\begin{equation}
    \chi = b {\rm{cosh}}^2 \qty(\frac{\phi}{2 \sqrt{a}})~,
\end{equation}
and thus the Lagrangian above becomes
\begin{equation}
    \mathcal{L}_{\chi} = - \frac{1}{2} \nabla_\mu \phi \nabla^\mu \phi
    - \frac{1}{\alpha_{\rm{R}}} \qty [b {\rm{cosh}}^2 \qty(\frac{\phi}{2 \sqrt{a}}) - c]^2~.
\end{equation}
This potential is possible to realize inflation if $c/b \gg 1$ \cite{Hong:2025tyi}, which approaches the limit of alpha-attractor inflation.
In the case of $a>0$ and $\chi <0$, one can obtain similar results by performing the field redefinition of $\chi = - b {\rm{sinh}}^2 [\phi/(2 \sqrt{a})]$.
Also, if $a<0$ and $0 < \chi < b$, one can use $\chi = b {\rm{sin}}^2 [\phi/(2 \sqrt{-a})]$ and choose suitable parameters to realize inflation (see \textit{e.g.} \cite{Boubekeur:2005zm}).

Another possibility is that the kinetic term has a local enhancement, whereas at the local maximum point the kinetic term has a finite value unlike the previous case.
Such a local enhancement can also lead to a flat potential which is suitable for inflation \cite{He:2025bli,Racioppi:2024zva,Racioppi:2024pno,Gialamas:2024uar} and can better fit the latest ACT data~\cite{ACT:2025fju,ACT:2025tim}.
We will give a specific example of this case in the next subsection.

\subsection{Example: Starobinsky inflation and its deformation with fermion chiral current coupling}
\label{sec:example1}

In this section, we consider some specific examples of the general results above.
We choose $S_\mu j^{\mu}_5$ as the coupling between the torsion components and the currents, where $j^{\mu}_5$ is the chiral current of a massive Dirac fermion.
We assume the fermion is in the fundamental representation of an $\mathrm{SU}(3)$ gauge group.
This particular setup is motivated by the claim that this coupling is related to the strong CP problem \cite{Mielke:2006zp,Mercuri:2009zi,Mercuri:2009zt,Lattanzi:2009mg,Castillo-Felisola:2015ema,Karananas:2018nrj,Karananas:2024xja}.
However, as we will see shortly, this is not the case; the coupling induced by this term alone has nothing to do with the strong CP problem.

Note that $S_\mu j^{\mu}_5$ naturally appears from the kinetic term of the fermion in EC gravity as illustrated in Sec.~\ref{sec:pre}.
For the geometric part, in the first example we use $\bar{R}$ of the EC gravity in the linear part and $\bar{R}$ and $\nabla_\mu S^\mu$
in the squared bracket in Eq.~\eqref{eq-general1}. 
It has been shown previously this setup can realize the $ \alpha $-attractor inflation \cite{Starobinsky:1980te,He:2024wqv,Ellis:2013nxa,Ferrara:2013rsa,Kallosh:2013yoa,Kallosh:2014rga,Carrasco:2015pla,Roest:2015qya,Linde:2015uga,Scalisi:2015qga}.
Another example is that we add only $\nabla_\mu S^\mu$ and $S_\mu T^\mu$ in the squared bracket while use $\bar{R}$ and $S_\mu T^\mu$ for the linear part, \textit{i.e.}, $\alpha = 0$.
This setup can realize deformations of the alpha-attractor inflation \cite{He:2024wqv,Salvio:2022suk}.
We specify the decay constants in the examples.

For the first example, we start from the action
\begin{align}
    \begin{split}
        S_\text{J} = \int \sqrt{-g_\text{J}} \dd^4 x\, & \left[
            \frac{\Mpl^2}{2} \qty(R_\text{J} + \frac{1}{24} S_{\mu} S^{\mu} - \frac{2}{3} T_{\mu} T^{\mu} )
            \right.\\
            & \left. 
            + \alpha_\text{R} \qty(R_\text{J} + \frac{1}{24} S_{\mu} S^{\mu} - \frac{2}{3} T_{\mu} T^{\mu} + \alpha_4 \nabla_\mu S^\mu + 2 \nabla_\mu T^\mu) ^ 2
            + \zeta S_\mu j^{\mu}_5 \right. \\
            &\left.
            + i \bar{\psi} \gamma^\mu 
            \qty(\partial_\mu - \frac{1}{8} \omega^{AB}_{\mu} [\gamma_A,\,\gamma_B] - i g A_\mu^a \tau^a) \psi
            - m \bar{\psi} \psi 
            - \frac{1}{2} {\rm{Tr}} F^{\mu \nu} F_{\mu \nu}\right]~,
    \end{split}
    \label{eq-defsu3}
\end{align}
where $\tau^a$s are generator matrices of SU(3), and $F_{\mu \nu} = \partial_\mu A_\nu - \partial_\nu A_\mu - i g [A_\mu, A_\nu] $ with $A_\mu \equiv A^a_\mu \tau^a$.
This is to take $\alpha = 1$, $\alpha_1 = \beta_1 = 1/24$, $\alpha_2 = \beta_2 = -2/3$, $ \beta_3 = 0$, $\alpha_5 = 2$ and $\alpha_4 \neq 0$ in Eq.~\eqref{eq-general1}.
The $S_\mu j^{\mu}_5$ coupling is $\zeta = 1/8$ if we only consider the contribution from the kinetic term of the fermion, while here we keep a general $\zeta$. 

Using the results in the previous subsection, the kinetic term in the Einstein frame is
\begin{equation}
    \mathcal{L}_{\text{kinetic}} = - 6 \alpha_4^2 \Mpl^2 \frac{\nabla_\mu \Omega^2 \nabla^\mu \Omega^2}{2 \Omega^4}~.
    \label{eq-eg1k}
\end{equation}
The canonical inflaton is thus
\begin{equation}
     \phi \equiv \sqrt{6} \alpha_4 \Mpl \text{ln} \Omega^2~.
     \label{eq-eg1phi}
\end{equation}
In this case,
\begin{equation}
    \mathcal{L}_{\text{E} 2} = 12 \alpha_4 \zeta \frac{\nabla_\mu \Omega^2}{\Omega^2} j^{\mu}_5
    = \frac{2 \sqrt{6} \zeta}{\Mpl} \nabla_\mu \phi j^{\mu}_5~,
    \label{eq-eg1E2}
\end{equation}
and the decay constant is independent of $ \phi $.
The self-coupling term of the current is
\begin{equation}
    \label{eq-eg1E3}
    \mathcal{L}_{\text{current}} = - \frac{12 \zeta^2}{\Mpl^2} j_{5\mu} j^{\mu}_5~,
\end{equation}
which together with Eq.~\eqref{eq-eg1E2} indicates that the cutoff of the effective field theory is $\sim \Mpl/|\zeta|$.
One can check the result of the decay constants matches the general asymptotic behavior shown in the last section. 
The current coupling~\eqref{eq-eg1E2} also provides a decay channel for the scalaron to produce chiral fermions~\cite{Kusenko:2014uta,Adshead:2015jza,DeSimone:2016ofp}. 
On top of this, the scalaron $\phi$ also couples to the fermion mass term through the conformal factor when going to the Einstein frame, which is
\begin{equation}
    \mathcal{L}_{\text{mass}} = - m \Omega \bar{\psi}_{\rm E} \psi_{\rm E} = -m \, e^{\frac{\phi}{2\sqrt{6} \alpha_4 \Mpl}} \bar{\psi}_{\rm E} \psi_{\rm E} ~, 
    \label{eq-eg1E4}
\end{equation}
where $ \psi_{\rm E} = \Omega^{-3/2} \psi $ is the canonicalized fermion in the Einstein frame. 
This interaction term allows a decay channel $ \phi \to \bar{\psi}_{\rm E} \psi_{\rm E} $ for the scalaron to decay into two fermions. 
This can be seen by expanding the exponent around the origin of $ \phi $, which gives a mass term for $ \psi $ as well as $ \propto m \,\phi \bar{\psi}_{\rm E} \psi_{\rm E} /(\alpha_4 \Mpl) $ at the first and second leading order.

Regarding the $\phi$ field as a constant background, one may readily examine whether the coupling in Eq.~\eqref{eq-eg1E2} leads to a non-trivial effective potential of $\phi$ after integrating out the QCD sector.
It is obvious that the current coupling in Eq.~\eqref{eq-eg1E2} never gives rise to the effective potential of $\phi$ because the current coupling in Eq.~\eqref{eq-eg1E2} vanishes for the constant background of $\phi$.
Thus this term has nothing to do with the strong CP problem (see \textit{e.g.} \cite{Kim:2008hd}).
Nevertheless, one might still wonder what if we perform the chiral rotation of the fermion field to make the current coupling vanish.
After the chiral rotation, we now have a Chern--Simons coupling and the chiral mass phase coupling, \textit{i.e.},
\begin{align}
    \frac{2\sqrt{6} \zeta\phi}{\Mpl}\frac{g^2}{8 \pi^2} \Tr F_{\mu\nu} \tilde F^{\mu\nu}
    - m \, e^{\frac{\phi}{2\sqrt{6} \alpha_4 \Mpl}} \bar\psi_{\rm E} e^{i \gamma_5 \frac{4 \sqrt{6} \zeta\phi}{\Mpl}} \psi_{\rm E}
    = \frac{2\sqrt{6} \zeta\phi}{\Mpl}\frac{g^2}{8 \pi^2} \Tr F_{\mu\nu} \tilde F^{\mu\nu}
    - i \frac{4 \sqrt{6} \zeta m}{\Mpl} \phi\bar\psi_{\rm E} \gamma_5 \psi_{\rm E} + \cdots
    ~,
\end{align}
where $\tilde F^{\mu\nu} = \frac{1}{2} \epsilon^{\mu\nu\rho\sigma} F_{\rho\sigma}$ is the dual field strength.
(see Appendix~\ref{app-rotation}). Again, one can readily check that the effective potential from the first term and the second in the RHS cancel each other in the constant background of $\phi$ as it should be because the physics does not depend on the choice of the field basis.
Note that the leading order contribution from ${\rm exp}(\phi/(2 \sqrt{6} \alpha_4 \Mpl))$ to the effective potential is due to (\ref{eq-eg1E4}) and is not related to the QCD $\theta$ term.
See also note added.

Now let us analyze another example.
The starting action is 
\begin{align}
    \begin{split}
        S = \int \sqrt{-g} \dd^4 x\, 
            &\left[\frac{\Mpl^2}{2} \qty(R + \frac{1}{24} S'_{\mu} S'^{\mu} - \frac{2}{3} T'_{\mu} T'^{\mu} + \beta_3' S'_\mu T'^\mu)
            + \alpha_\text{R} \qty(\nabla_\mu S'^\mu + \alpha_3' S'_\mu T'^\mu) ^ 2
            + \zeta' S'_\mu j^{\mu}_5 \right. \\
            &\left.
            + i \bar{\psi} \gamma^\mu 
            \qty(\partial_\mu - \frac{1}{8} \omega^{AB}_{\mu} [\gamma_A,\,\gamma_B] - i g A_\mu^a T^a) \psi
            - m \bar{\psi} \psi 
            - \frac{1}{2} {\rm{Tr}} F^{\mu \nu} F_{\mu \nu}
            \right]~,
    \end{split}
\end{align}
with $\alpha_3' \neq 0$.
The primes mean that we have not yet done field redefinition to remove the redundancy of $S_\mu T^\mu$ as mentioned in the previous subsection. 
Doing Legendre transformation and solving constraint equations for torsion components, one finds the kinetic term as
\begin{equation}
    \mathcal{L}_{\text{kinetic}} = -\frac{96 \Mpl^2}{\Mpl^4 + 9 (\beta_3' \Mpl^2 + 4 \alpha_3' \chi)^2} \frac{\nabla^\mu \chi \nabla_\mu \chi}{2}~.
\end{equation}
Setting
\begin{equation}
    \chi = \frac{\Mpl^2}{12 \alpha_3'} \qty[ \text{sinh} \qty( \pm \sqrt{\frac{3}{2}} \frac{\alpha_3'}{\Mpl} \phi) - 3 \beta_3']~,
\end{equation}
one obtains a canonicalized kinetic term with $\phi$ as the inflaton.
This is an example of the regularized pole inflation \cite{He:2024wqv,He:2025bli} as mentioned in last subsection.
We take the minus sign in the hyperbolic sine without losing generality because the field redefinition $\phi \mapsto -\phi$ is possible.

The coupling between the inflaton and the current in this model is
\begin{equation}
    \mathcal{L}_{2} = \frac{48 \Mpl^2 \zeta \nabla_\mu \chi}{\Mpl^4 + 9 (\beta_3' \Mpl^2 + 4 \alpha_3' \chi) ^ 2} j^{\mu}_5
    = - \frac{2 \sqrt{6}}{\Mpl \text{cosh} \qty(\sqrt{\frac{3}{2}} \frac{\alpha_3'}{\Mpl} \phi)} \zeta \nabla_\mu \phi j^{\mu}_5.
    \label{eq-ex1}
\end{equation}
For the same reason as the previous example, this coupling alone has nothing to do with the strong CP problem since the effective potential vanishes for a constant background of $\phi$ after integrating out the QCD sector.
Also, the self-coupling term of currents is
\begin{equation}
    \mathcal{L}_{3} = - \frac{12 \zeta^2}{\Mpl^2 \text{cosh}^2 \qty(\sqrt{\frac{3}{2}} \frac{\alpha_3'}{\Mpl} \phi)} j^{\mu}_5 j_{5 \mu} ~.
\end{equation}

To compare with the general results of the behavior of the decay constants shown above, one can redefine $S^{\mu} = S'^\mu + T'^\mu$ and $T^{\mu} = - S'^\mu + T'^\mu$. 
The new sets of coefficients are thus
\begin{equation}
    \beta_1 = \frac{1}{4} \qty(- \frac{5}{8} + \beta_3')~, \qquad
    \beta_2 = -\frac{1}{4} \qty( \frac{5}{8} + \beta_3')~, \qquad
    \beta_3 = - \frac{17}{48}~,
\end{equation}
\begin{equation}
    \alpha_1 = \frac{\alpha_3'}{4}~, \qquad
    \alpha_2 = - \frac{\alpha_3'}{4}~, \qquad
    \alpha_4 = \frac{1}{2}~, \qquad
    \alpha_5 = - \frac{1}{2}~, \qquad
    \zeta = \frac{\zeta'}{2}~, \qquad
    \xi = - \frac{\zeta'}{2}~.
    \label{eq-ex2}
\end{equation}
Note that in this case, $4 \beta_1 \beta_2 - \beta_3^2 < 0$, $\alpha_2 \alpha_4^2 + \alpha_1 \alpha_5^2 = 0$ and $\alpha_2 \alpha_4 \zeta + \alpha_1 \alpha_5 \xi = 0$. 
The asymptotic behavior can thus be found in Table~\ref{table-inf} and Table~\ref{table-0}, which match Eqs.~\eqref{eq-ex1} and \eqref{eq-ex2}.

We briefly comment on the reheating process after inflation in this setup, especially in the presence of vector/chiral current couplings. 
Through conformal coupling, the scalaron can couple with the mass term of the fermions, which will lead to a decay channel of the scalaron $ \phi \to \bar{\psi} \psi $, similar to the Starobinsky model. 
This coupling is Planck-suppressed since it is originated from gravity, so the efficiency is expected to be small $ \propto m_{\phi} \, (m/\Mpl)^2 $. 
Besides, the coupling between scalaron and vector/chiral currents can also lead to fermion production, which has been studied in Refs.~\cite{Salvio:2022suk,Dolgov:1994zq,Adshead:2015kza,Domcke:2018eki}. 
The perturbative decay rate of the scalaron via this operator also scales as $m_\phi (m / \Mpl)^2$. 
On the other hand, the motion of the scalaron can act as an effective chemical potential for the (chiral) fermions, which can be non-trivial if the scalaron motion does not cancel out the effects by itself, for example, via oscillation. 
This may be interesting for topics like baryogenesis~\cite{Kusenko:2014uta,Adshead:2015jza,DeSimone:2016ofp} and chiral gravitational wave production~\cite{Anber:2016yqr}.

\section{Gauge-dependent currents}
\label{sec:gauge-dependent}

In the previous section, we have seen that the chiral current coupling can be naturally induced in EC gravity, and also confirmed that this coupling alone has nothing to do with the strong CP problem.
In this section, we extend the construction of the previous section so as to include the Chern--Simons coupling in this framework.
As we will clarify later, this coupling is indeed related to the $\theta$ term of the QCD sector.
However, in order for this scalaron to be a viable candidate of QCD axion, we need to make sure that there is no other contribution to the effective potential of the scalaron other than the QCD sector, which is generically not the case in the simple setups.

Specifically, the gauge-dependent currents under consideration in this section have the following property that is the same as the Chern--Simons current; while the current itself depends on the gauge transformation, its total derivative is gauge invariant, \textit{i.e.,}
\begin{equation}
    \label{eq-gauge-dep-current}
    \star j \mapsto \star j + \dd \omega_2 (\theta, A) \quad \Rightarrow \quad \dd \star j \mapsto \dd \star j ~,
\end{equation}
under the gauge transformation of the gauge field $A \mapsto A + \dd \theta / g$.
Here the current is denoted as a one-form $j = j^\mu \dd x_\mu$ and the Hodge dual is denoted by the star operator $\star$.
$\omega_2$ is a two-form depending on the gauge transformation parameter $\theta$ and the gauge field $A$.
The central question here is how to couple gauge-dependent currents to the torsion components.
The action becomes gauge-dependent if some of the currents in Eq.~\eqref{eq-general1} are gauge-dependent currents such as the Chern--Simons current.

However, if we couple the currents to the torsion components in the following way:
\begin{align}
    \begin{split}
        S_\text{J} = \int \sqrt{-g_\text{J}} \dd^4 x\, & \Bigg\{
            \frac{\Mpl^2}{2} \left[R_\text{J} 
            + \beta_1 \qty( S^{\mu} + \sum_n \frac{\zeta_n}{\Mpl^2} j^\mu_n)^2 
            + \beta_2 \qty( T^{\mu} + \sum_n \frac{\xi_n}{\Mpl^2} j^\mu_n)^2
            \right.\\
            & \left.
            + \beta_3 \qty( S^{\mu} + \sum_n \frac{\zeta_n}{\Mpl^2} j^\mu_n) \qty( T^{\mu} + \sum_n \frac{\xi_n}{\Mpl^2} j^\mu_n)\right] \\
            & 
            + \alpha_\text{R} \left[\alpha R_\text{J} 
            + \alpha_1 \qty( S^{\mu} + \sum_n \frac{\zeta_n}{\Mpl^2} j^\mu_n)^2 
            + \alpha_2 \qty( T^{\mu} + \sum_n \frac{\xi_n}{\Mpl^2} j^\mu_n)^2 \right.\\
            & \left.
            + \alpha_3 \qty( S^{\mu} + \sum_n \frac{\zeta_n}{\Mpl^2} j^\mu_n) \qty( T^{\mu} + \sum_n \frac{\xi_n}{\Mpl^2} j^\mu_n) + \alpha_4 \nabla_\mu S^\mu + \alpha_5 \nabla_\mu T^\mu \right] ^ 2
            \Bigg\}~,
    \end{split}
\end{align}
the action is gauge-invariant.
This is because the gauge dependent parameter $\omega_2$ is absorbed by the following simultaneous redefinition of the torsion components:
\begin{align}
    \star S &\mapsto \star S - \sum_n \frac{\zeta_n}{\Mpl} \dd \omega_{n2} (\theta, A) \quad \Rightarrow \quad \dd \star S \mapsto \dd \star S ~, \\
    \star T & \mapsto \star T - \sum_n \frac{\xi_n}{\Mpl} \dd \omega_{n2} (\theta, A) \quad \Rightarrow \quad \dd \star T \mapsto \dd \star T ~.
\end{align}
One may immediately see that these transformations never change the total derivative of $T_\mu$ and $S_\mu$ and hence leave the action invariant.
Here we again use the form notation for brevity, \textit{i.e.}, $S = S_\mu \dd x^\mu$ and $T = T_\mu \dd x^\mu$.

One can make this property more explicit by performing the following field redefinition:
\begin{equation}
S'^\mu \equiv S^\mu + \sum_n \frac{\zeta_n}{\Mpl^2} j^\mu_n, \quad
T'^\mu \equiv T^\mu + \sum_n \frac{\xi_n}{\Mpl^2} j^\mu_n~, 
\end{equation}
and replacing the notation of $S'^\mu$ and $T'^\mu$ with $S^\mu$ and $T^\mu$, respectively. 
Then the action above becomes
\begin{align}
    \label{eq-gi}
    \begin{split}
        S_\text{J} = \int \sqrt{-g_\text{J}} \dd^4 x\, & \Bigg\{
            \frac{\Mpl^2}{2} \qty(R_\text{J} + \beta_1 S_{\mu} S^{\mu} + \beta_2 T_{\mu} T^{\mu} + \beta_3 S_{\mu} T^{\mu} )
            \\
            & 
            + \alpha_\text{R} \Bigg[\alpha R_\text{J} + \alpha_1 S_{\mu} S^{\mu} + \alpha_2 T_{\mu} T^{\mu} + \alpha_3 S_{\mu} T^{\mu} 
            \\
            &
             + \alpha_4 \nabla_\mu \qty(S^\mu - \sum_n \frac{\zeta_n}{\Mpl^2} j^\mu_n) + \alpha_5 \nabla_\mu \qty(T_\mu - \sum_n \frac{\xi_n}{\Mpl^2} j^\mu_n) \Bigg] ^ 2
            \Bigg\}~.
    \end{split}
\end{align}
This action is obviously gauge-invariant since $\nabla_\mu j_n^\mu$ is gauge-invariant by definition \eqref{eq-gauge-dep-current}.
Applying this formulation to the Chern--Simons current, we expect that the scalar degree of freedom obtains the Chern--Simons coupling, $\phi F \tilde F$, different from the results in last section, which only gives $\nabla\phi j_5$-type couplings.
Contrary to the current coupling $\nabla\phi j_5$, the Chern--Simons coupling $\phi F \tilde F$ cannot be rotated away for a constant background of $\phi$ as long as there exists the strong CP problem, \textit{i.e.,} the $\theta$ term is physical.\footnote{
    If massless fermions are involved, the $\theta$ term is unphysical since this term can be rotated away to be $\nabla \theta j_5$ that is obviously vanishing as $\theta$ is a constant.
}
The effective potential of $\phi$ receives the periodic contribution with respect to $\theta + \zeta \phi / \Mpl$ with $\zeta$ being the parameter of the coupling.
Therefore, the current coupling introduced in this section is indeed related to the $\theta$ term in the QCD sector.
Although it can in principle serve as the QCD axion as long as the scalar degree of freedom $\phi$ has a shift symmetry when the Chern--Simons coupling is turned off, this is generically not the case as this scalaron usually couples to the dimensionful parameter such as the Planck mass, Higgs mass, or the running of the couplings as will be explained later.

We further add gauge-independent currents to the linear part of Eq.~\eqref{eq-gi} as in the previous section.
The general results are given in Sec.~\ref{sec:general2}, and two examples are given in Sec.~\ref{sec:example2}.

\subsection{General results}
\label{sec:general2}
We consider the following as the starting action, 
\begin{align}
    \label{eq-gi-1}
    \begin{split}
        S_\text{J} = \int \sqrt{-g_\text{J}} \dd^4 x\, & \Bigg\{
            \frac{\Mpl^2}{2} \qty(R_\text{J} + \beta_1 S_{\mu} S^{\mu} + \beta_2 T_{\mu} T^{\mu} + \beta_3 S_{\mu} T^{\mu} )
            + \sum_n \qty(\zeta_{1n} S_\mu j_n^{\rm{gi}\mu} + \xi_{1n} T_\mu j_n^{\rm{gi}\mu})
            \\
            & 
            + \alpha_\text{R} \Bigg[ \alpha R_\text{J} + \alpha_1 S_{\mu} S^{\mu} + \alpha_2 T_{\mu} T^{\mu} + \alpha_3 S_{\mu} T^{\mu} 
            \\
            &
             + \alpha_4 \nabla_\mu \qty(S^\mu - \sum_n \frac{\zeta_{2n}}{\Mpl^2} j^{\rm{gi}\mu}_n -\sum_m \frac{\zeta_{3m}}{\Mpl^2} j^{\rm{gd}\mu}_m) + \alpha_5 \nabla_\mu \qty(T^\mu - \sum_n \frac{\xi_{2n}}{\Mpl^2} j^{\rm{gi}\mu}_n -\sum_m \frac{\xi_{3m}}{\Mpl^2} j^{\rm{gd}\mu}_m) \Bigg] ^ 2
            \Bigg\}~,
    \end{split}
\end{align}
where $n$ sums over gauge-independent currents and $m$ sums over gauge-dependent currents, and $\alpha_\text{R} \neq 0$.

From now on we set $\alpha_3 = 0$ as in the previous section to reduce redundancy.
Legendre transformation leads to
\begin{align}
    \begin{split}
        S_\text{J} = \int \sqrt{-g_\text{J}} \dd^4 x\, & \left\{
            \frac{\Mpl^2}{2} \qty(R_\text{J} + \beta_1 S_{\mu} S^{\mu} + \beta_2 T_{\mu} T^{\mu} + \beta_3 S_{\mu} T^{\mu} )+ \sum_n \qty(\zeta_{1n} S_\mu j_n^{\rm{gi}\mu} + \xi_{1n} T_\mu j_n^{\rm{gi}\mu})
            \right.\\
            & 
            + \Bigg[\alpha R_\text{J} + \alpha_1 S_{\mu} S^{\mu} + \alpha_2 T_{\mu} T^{\mu}
            \\
            & 
             + \alpha_4 \nabla_\mu \qty(S^\mu - \sum_n \frac{\zeta_{2n}}{\Mpl^2} j^{\rm{gi}\mu}_n -\sum_m \frac{\zeta_{3m}}{\Mpl^2} j^{\rm{gd}\mu}_m) \\
             & \left. + \alpha_5 \nabla_\mu \qty(T^\mu - \sum_n \frac{\xi_{2n}}{\Mpl^2} j^{\rm{gi}\mu}_n -\sum_m \frac{\xi_{3m}}{\Mpl^2} j^{\rm{gd}\mu}_m) \Bigg] 2 \chi - \frac{\chi^2}{\alpha_\text{R}} 
            \right\}~.
    \end{split}
\end{align}
We then solve the constraint equations of $S_\mu$ and $T_\mu$ and obtain
\begin{align}
    \begin{split}
        S_\text{J} = \int \sqrt{-g_\text{J}} \dd^4 x\, & \Bigg\{
            \frac{\Mpl^2}{2} \Omega^2 R_\text{J} -\frac{\chi ^ 2}{\alpha_\text{R}}
            \\
            & 
            - \left[
             \qty(\frac{\Mpl ^ 2}{2} \beta_2 + 2 \alpha_2 \chi) \qty( \sum_n \zeta_{1n} j_{n \mu}^{\rm{gi}} - 2 \alpha_4 \nabla_\mu \chi ) ^2 
            \right.\\
            &
            - \frac{\Mpl ^ 2}{2} \beta_3 \qty( \sum_n \zeta_{1n} j_{n \mu}^{\rm{gi}} - 2 \alpha_4 \nabla_\mu \chi ) \qty(\sum_n \xi_{1n} j_{n}^{\rm{gi}\mu} - 2 \alpha_5 \nabla^\mu \chi)\\
            & \left.
            + \qty(\frac{\Mpl ^ 2}{2} \beta_1 + 2 \alpha_1 \chi) \qty( \sum_n \xi_{1n} j_{n \mu}^{\rm{gi}} - 2 \alpha_5 \nabla_\mu \chi ) ^2 \right] \\
            & \times
            \qty[4 \qty(\frac{\Mpl ^ 2}{2} \beta_1 + 2 \alpha_1 \chi) \qty(\frac{\Mpl ^ 2}{2} \beta_2 + 2 \alpha_2 \chi) - \qty(\frac{\Mpl ^ 2}{2} \beta_3 ) ^ 2]^{-1} \\
            & - 2 \alpha_4 \chi \nabla_\mu \qty(\sum_n \frac{\zeta_{2n}}{\Mpl^2} j^{\rm{gi}\mu}_n +\sum_m \frac{\zeta_{3m}}{\Mpl^2} j^{\rm{gd}\mu}_m) - 2 \alpha_5 \chi \nabla_\mu \qty(\sum_n \frac{\xi_{2n}}{\Mpl^2} j^{\rm{gi}\mu}_n + \sum_m \frac{\xi_{3m}}{\Mpl^2} j^{\rm{gd}\mu}_m)
            \Bigg\}~,
    \label{eq-gdgeneral}
    \end{split}
\end{align}
where $\Omega^2 = 1 + 4 \alpha \chi / \Mpl^2 > 0$.
Note that except for the last line, this action is the same as Eq.~\eqref{eq-action1}.
The discussion of inflation possibilities in Sec.~\ref{sec:noasymp} section therefore also applies here.

We now focus on the last line of Eq.~\eqref{eq-gdgeneral}:
\begin{equation}
\mathcal{L}_{\text{J} 3} = - 2 \alpha_4 \chi \nabla_\mu \qty(\sum_n \frac{\zeta_{2n}}{\Mpl^2} j^{\rm{gi}\mu}_n +\sum_m \frac{\zeta_{3m}}{\Mpl^2} j^{\rm{gd}\mu}_m) - 2 \alpha_5 \chi \nabla_\mu \qty(\sum_n \frac{\xi_{2n}}{\Mpl^2} j^{\rm{gi}\mu}_n + \sum_m \frac{\xi_{3m}}{\Mpl^2} j^{\rm{gd}\mu}_m)~.
\end{equation}
According to Eq.~\eqref{eq-conformal2}, after Wely transformation, this part has the same form:
\begin{equation}
\mathcal{L}_{\text{E} 3} = - 2 \alpha_4 \chi \nabla_\mu \qty(\sum_n \frac{\zeta_{2n}}{\Mpl^2} j^{\rm{gi}\mu}_n +\sum_m \frac{\zeta_{3m}}{\Mpl^2} j^{\rm{gd}\mu}_m) - 2 \alpha_5 \chi \nabla_\mu \qty(\sum_n \frac{\xi_{2n}}{\Mpl^2} j^{\rm{gi}\mu}_n + \sum_m \frac{\xi_{3m}}{\Mpl^2} j^{\rm{gd}\mu}_m)~,
\label{eq-gdE3}
\end{equation}
with currents defined in the Einstein frame but we have omitted the subscript ``E'' for convenience.
Integrating by part and using the canonicalized inflaton Eq.~\eqref{eq-cano}, one can rewrite it as
\begin{align}
    \begin{split}
        \mathcal{L}_{\text{E} 3} =
        & \frac{1}{\sqrt{\frac{24 \alpha^2}{\Omega^4 \Mpl^2} + \frac{F(\chi(\phi))}{\Omega^2}}} \nabla_\mu \phi \qty(2 \alpha_4 \sum_n \frac{\zeta_{2n}}{\Mpl^2} j^{\rm{gi}\mu}_n + 2 \alpha_5 \sum_n \frac{\xi_{2n}}{\Mpl^2} j^{\rm{gi}\mu}_n) \\
        & - 2 \alpha_4 \chi(\phi) \sum_m \frac{\zeta_{3m}}{\Mpl^2} \nabla_\mu j^{\rm{gd}\mu}_m
        - 2 \alpha_5 \chi(\phi) \sum_m \frac{\xi_{3m}}{\Mpl^2} \nabla_\mu j^{\rm{gd}\mu}_m ~.
    \end{split}
    \label{eq-gauge-inv-einstein}
\end{align}
This part has a relatively simple form of decay constant compared to the cases discussed in the previous section.
In this way we are able to obtain the couplings between the inflaton and the gauge-dependent currents. 
Note that the second line of Eq.~\eqref{eq-gauge-inv-einstein} can give $\phi F \tilde{F}$-type couplings directly.
We list the asymptotic behavior of the coefficients of these terms in Tables~\ref{table-inf-gd-1}-\ref{table-inf-gd-4}. 
The discussion of the cutoff of EFT at the end of Sec.~\ref{sec:asymp} also applies here.

\NiceMatrixOptions{cell-space-limits=4pt}
\begin{table}
    \caption{Behavior of coefficients in Eq.~\eqref{eq-gauge-inv-einstein} at $\chi \to \infty$ for $\alpha = 0$.}
\begin{center}
\begin{NiceTabular}{|c|c|c|}\hline
    $F(\chi) \sim $ & $ 1/\sqrt{\frac{24\alpha^2}{\Omega^4 \Mpl^2} + \frac{F}{\Omega^2}} \sim$ & $ \chi(\phi) \sim $ \\ \hline

    $ \chi $ & $ \phi^{-\frac{1}{3}} $ & $ \phi^{\frac{2}{3}} $ \\ \hline
    
    const. & const. & $ \phi $ \\ \hline

    $ \chi^{-1} $ & $ \phi $ & $ \phi^2 $ \\ \hline

    $ \chi^{-2} $ & $ e^\phi $ & $ e^\phi $ \\ \hline

\end{NiceTabular}
\end{center}
\label{table-inf-gd-1}
\end{table}
\NiceMatrixOptions{cell-space-limits=4pt}
\begin{table}
    \caption{Behavior of coefficients in Eq.~\eqref{eq-gauge-inv-einstein} at $\chi \to \infty$ for $\alpha \neq 0$.}
\begin{center}
\begin{NiceTabular}{|c|c|c|}\hline
    $F(\chi) \sim $ & $ 1/\sqrt{\frac{24\alpha^2}{\Omega^4 \Mpl^2} + \frac{F}{\Omega^2}} \sim$ & $ \chi(\phi) \sim $ \\ \hline

    $ \chi $ & const. & $ \phi $ \\ \hline

    const. & $ \phi $ & $ \phi^2 $ \\ \hline

    $ \chi^{-1} $ & $ e^\phi $ & $ e^\phi $ \\ \hline

    $ \chi^{-2} $ & $ e^\phi $ & $ e^\phi $ \\ \hline

\end{NiceTabular}
\end{center}
\label{table-inf-gd-2}
\end{table}
\NiceMatrixOptions{cell-space-limits=4pt}
\begin{table}
    \caption{Behavior of coefficients in Eq.~\eqref{eq-gauge-inv-einstein} at $\chi \to 0 $ for $\alpha = 0$.}
\begin{center}
\begin{NiceTabular}{|c|c|c|}\hline
    $F(\chi) \sim $ & $ 1/\sqrt{\frac{24\alpha^2}{\Omega^4 \Mpl^2} + \frac{F}{\Omega^2}} \sim$ & $ \chi(\phi) \sim $ \\ \hline

    $ \chi $ & $ \phi^{-\frac{1}{3}} $ & $ \phi^{\frac{2}{3}} $ \\ \hline

    const. & const. & $ \phi $ \\ \hline

    $ \chi^{-1} $ & $ \phi $ & $ \phi^2 $ \\ \hline
    
    $ \chi^{-2} $ & $ e^\phi $ & $ e^\phi $ \\ \hline

\end{NiceTabular}
\end{center}
\label{table-inf-gd-3}
\end{table}
\NiceMatrixOptions{cell-space-limits=4pt}
\begin{table}
    \caption{Behavior of coefficients in Eq.~\eqref{eq-gauge-inv-einstein} at $\chi \to 0 $ for $\alpha \neq 0$.}
\begin{center}
\begin{NiceTabular}{|c|c|c|}\hline
    $F(\chi) \sim $ & $ 1/\sqrt{\frac{24\alpha^2}{\Omega^4 \Mpl^2} + \frac{F}{\Omega^2}} \sim$ & $ \chi(\phi) \sim $ \\ \hline

    $ \chi $ & const. & $ \phi $ \\ \hline

    const. & const. & $ \phi $ \\ \hline

    $ \chi^{-1} $ & $ \phi $ & $ \phi^2 $ \\ \hline
    
    $ \chi^{-2} $ & $ e^\phi $ & $ e^\phi $ \\ \hline

\end{NiceTabular}
\end{center}
\label{table-inf-gd-4}
\end{table}

\subsection{Examples: Chern--Simons current couplings}
\label{sec:example2}

In this section, we consider two examples.

In Sec.~\ref{sec:CP}, as a counterpart of the first example in Sec.~\ref{sec:example1}, we provide an explicit example where the scalar degree of freedom has the Chern--Simons coupling to $\mathrm{SU}(3)$ gauge field.
We also add a massive Dirac fermion charged under this gauge group in the same way as in Sec.~\ref{sec:example1}.
This example clearly demonstrates that the scalar degree of freedom receives the periodic potential from the QCD sector in a particular combination of the QCD $ \theta $ angle and the Chern--Simons coupling parameter.
Nevertheless, the scalaron does not serve as QCD axion generically because the scalaron usually couples to the dimensionful parameters such as the Planck mass, Higgs mass, or the running of the couplings, which gives rise to the effective potential with its minimum different from the QCD $\theta$ term.

In Sec.~\ref{sec:gauge}, we instead discuss the effect of these couplings when the scalar degree of freedom is identified as inflaton.
It is well known that the Chern--Simons coupling gives rise to the efficient gauge field production even during inflation.
For an illustrative purpose, we decouple the fermion, and focus on the implications of the Chern--Simons coupling to the inflationary dynamics.

\subsubsection{Necessary condition of strong CP problem}
\label{sec:CP}
We introduce $j_1^{{\rm{gd}}\mu}=E^{\mu\nu\lambda\rho}{\rm{Tr}}(F_{\nu\lambda}A_\rho-2/3\,A_\nu A_\lambda A_\rho)$ with $A_\mu$ and $F_{\mu\nu}$ defined below Eq.~\eqref{eq-defsu3}, which is the Chern--Simons current of SU(3) gauge field, and $j_1^{{\rm{gi}}\mu}= \bar{\psi} \gamma^\mu \gamma_5 \psi \equiv j^{\mu}_5$, which is the chiral current of a massive fermion.
We set $\xi_{11} = \xi_{21} = \xi_{31} = 0$ for simplicity in this example.
For the geometry part, we consider the $ \alpha $-attractor inflation \cite{Starobinsky:1980te,He:2024wqv,Ellis:2013nxa,Ferrara:2013rsa,Kallosh:2013yoa,Kallosh:2014rga,Carrasco:2015pla,Roest:2015qya,Linde:2015uga,Scalisi:2015qga}, \textit{i.e.}, the first example in the previous section.

The starting action is then
\begin{align}
    \begin{split}
        S_\text{J} = \int \sqrt{-g_\text{J}} \dd^4 x\, & \left\{
            \frac{\Mpl^2}{2} \qty(R_\text{J} + \frac{1}{24} S_{\mu} S^{\mu} - \frac{2}{3} T_{\mu} T^{\mu} ) + \zeta_{11} S_\mu j_1^{\rm{gi}\mu}
            \right.\\
            & 
            + \alpha_\text{R} \qty[R_\text{J} + \frac{1}{24} S_{\mu} S^{\mu} - \frac{2}{3} T_{\mu} T^{\mu} + \alpha_4 \nabla_\mu \qty(S^\mu - \frac{\zeta_{21}}{\Mpl^2} j_1^{\rm{gi}\mu} - \frac{\zeta_{31}}{\Mpl^2} j_1^{\rm{gd}\mu}) + 2 \nabla_\mu T_\mu] ^ 2 \\
            &\left.
            + i \bar{\psi} \gamma^\mu 
            \qty(\partial_\mu - \frac{1}{8} \omega^{AB}_{\mu} [\gamma_A,\,\gamma_B] - i g A_\mu^a T^a) \psi
            - m \bar{\psi} \psi 
            - \frac{1}{2} {\rm{Tr}} F^{\mu \nu} F_{\mu \nu}
            \right\}~,
    \end{split}
\end{align}
The inflaton coupling to the chiral current $j_1^{\rm{gi}\mu}$ from the contribution of $\zeta_{11}$ term is the same as in Eq.~\eqref{eq-eg1E2}:
\begin{equation}
    \mathcal{L}_{\text{E} 2} 
    = \frac{2 \sqrt{6} \zeta_{11}}{\Mpl} \nabla_\mu \phi j_1^{\rm{gi}\mu}~.
\end{equation}
From Eq.~\eqref{eq-gdE3}, one can show that in this example, after dropping boundary term of the gauge-independent current,
\begin{align}
    \mathcal{L}_{\text{E} 3} 
    = -\frac{\alpha_4}{2}
    e^{\frac{\phi}{\sqrt{6} \alpha_4 \Mpl}}
    \zeta_{21} \nabla_\mu j_1^{\rm{gi} \mu} - \frac{\alpha_4}{2} \left( e^{\frac{\phi}{\sqrt{6} \alpha_4 \Mpl}} -1 \right) \zeta_{31} \nabla_\mu j_1^{\rm{gd} \mu} ~.
\end{align}
These terms can be written explicitly as
\begin{align}
    S_{\rm{E}2 + \rm{E}3} = \int \sqrt{-g_\text{E}} \dd^4 x\,
    & \left[
    \qty(\frac{2 \sqrt{6} \zeta_{11}}{\Mpl} \nabla_\mu \phi + \frac{\alpha_4 \zeta_{21}}{2} \nabla_\mu e^{\frac{\phi}{\sqrt{6} \alpha_4 \Mpl}}) j^{\mu}_5 
    -\frac{\alpha_4}{2} \left(
    e^{\frac{\phi}{\sqrt{6} \alpha_4 \Mpl}} -1 \right)
    \frac{\zeta_{31}}{2} {\rm{Tr}} E^{\mu \nu \alpha \beta} F_{\mu \nu} F_{\alpha \beta}
    \right]~.
    \label{eq-resultCP}
\end{align}
It is clear now that we added $\phi\nabla j^\mu_5$-type terms and $\phi F \tilde{F}$-type terms separately in the Lagrangian.
By integrating out the QCD sector under the background of the scalaron $\phi$, one may obtain the axion-type potential by non-perturbative effects of the QCD sector.
However, the scalaron is generically not a viable candidate for QCD axion because it usually couples to the dimensionful parameters such as the Planck mass, Higgs mass, or the running of the couplings.
One may kill the tree level potential from $\Mpl$ by taking the limit of $\alpha_\mathrm{R} \to \infty$, still, the scalaron in general couples to the other dimensionful parameters in the theory such as the Higgs mass or the running of the couplings.
One may resolve this issue by completely decoupling the scalaron from the dimensionful parameters \textit{via} the no-scale construction \cite{Shaposhnikov:2008xi,Karananas:2025ews,Hong:2025tyi}, which is beyond the scope of this paper.

\subsubsection{Gauge field production}
\label{sec:gauge}

We apply the general Lagrangian to gauge field production during inflation.
For simplicity, instead of SU(3) gauge field we consider U(1) gauge field.
The gauge-dependent Chern--Simons current of the gauge field is then $j_1^{\rm{gd}\mu} = E^{\mu \nu \lambda \rho} F_{\nu \lambda} A_\rho$ with $ A_\mu $ the U(1) gauge field and $F_{\nu \lambda} \equiv \partial_\nu A_\lambda - \partial_\lambda A_\nu$.
We also add the gravitational Chern--Simons current $j_2^{\rm{gd}\mu} =  2 E^{\mu \nu \lambda \rho} \omega_\nu{}^{IJ} (\partial_\lambda \omega_{\rho IJ} + 2 \omega_{\lambda I}{}^{K} \omega_{\rho KJ} / 3)$. $j_1^{\rm{gi}\mu} = \bar{\psi} \gamma^\mu \gamma_5 \psi \equiv j^{\mu}_5$ as the chiral current of a massive Dirac fermion at curved spacetime charged under the gauged U(1).
We assume the mass is so large that the fermion decouples.
We also set $\xi_{11} = \xi_{21} = \xi_{31} = \xi_{32} = 0$ for simplicity in this example and consider the alpha-attractor inflation.

Compared with Eq.~(\ref{eq-resultCP}) in the last subsection, we drop the terms with $j^\mu_5$ since we assume the fermion has mass much larger than the inflation scale and thus is decoupled.
Noting that we add the Chern--Simons current of gravity, we then have
\begin{align}
    S_{\rm{E}2 + \rm{E}3} = \int \sqrt{-g_\text{E}} \dd^4 x\,
    \left[
    -\frac{\alpha_4}{2}
    \left( e^{\frac{\phi}{\sqrt{6} \alpha_4 \Mpl}} -1 \right)
    \qty(\frac{\zeta_{31}}{2} E^{\mu \nu \alpha \beta} F_{\mu \nu} F_{\alpha \beta}
    + 
    \frac{\zeta_{32}}{2} E^{\mu \nu \alpha \beta} R_{\mu \nu A B} R_{\alpha \beta}{}^{A B} )
    \right]~.
\end{align}
Now we want to constrain the parameter range in this model by requiring that there are no ghost instabilities of chiral gravitational waves and no large non-Gaussianities from gauge field production. 

First we consider constraint that the ghost instabilities of the tensor modes are absent.
According to Eq.~\eqref{eq-ghost}, the physical momentum $k/a$ should satisfy
\begin{equation}
    \frac{k}{a} < \frac{\Mpl^2}{8 |\dot{f}|}
    \label{eq-ghost2} ~,
\end{equation}
to avoid ghost instability, where
\begin{equation}
    f \equiv 
     - \zeta_{32} \frac{\alpha_4}{2} \left(
    e^{\frac{\phi}{\sqrt{6} \alpha_4 \Mpl}} - 1 \right) ~,
\end{equation}
and the dot denotes derivative respect to physical time $t$
(see Appendix~\ref{app:chiralGW}). 
One may require the right-hand side of Eq.~\eqref{eq-ghost2} to be larger than $\Mpl$.
One may end up with a wider parameter range setting it to be only larger than the energy scale of the inflation $\sim H$, since the typical physical momentum should be lower than that value.
In other words, for the analysis to be valid for the process concerned, \textit{i.e.}, inflation and reheating, during such process, the following inequality must be satisfied:
\begin{equation}
\left|\dot{f} \right| < \frac{\Mpl}{8}
~~
\text{or}
~~
\left|\dot{f}\right| < \frac{\Mpl^2}{8H}~,
\end{equation}
where 
\begin{equation}
    \dot{f} = -\frac{1}{\sqrt{6} \Mpl}  \frac{\zeta_{32}}{2} {e}^{\frac{\phi}{\sqrt{6} \alpha_4 \Mpl}} \dot{\phi}~.
\end{equation}
From now on we set $\alpha_4>0$\footnote{One can do field redefinition $\phi \to -\phi$ for $\alpha_4 > 0$. }. 
During inflation, we can approximate $\dot{\phi}$ as
\begin{align}
    \dot{\phi}  = - \Mpl H \sqrt{2 \epsilon} 
     \approx  - \Mpl^2 H \left| \frac{V,_{\phi}}{V} \right| 
    &= - \Mpl H \frac{2}{\sqrt{6} \alpha_4} \frac{1}{e^{\frac{\phi}{\sqrt{6} \alpha_4 \Mpl}} - 1}~, 
\end{align}
where $\epsilon$ is the first slow-roll parameter, and $\alpha_4 \lesssim 1$ to match the observation.
Here we used the inflation potential:
\begin{equation}
V(\phi) = \frac{\Mpl^4}{16 \alpha_{\rm{R}}} \qty[1 - {\rm{exp}}\qty({-\frac{\phi}{\sqrt{6} \alpha_4 \Mpl}})]^2~. 
\end{equation}
At the end of inflation, $\epsilon = 1$ 
which gives $ \phi_{\rm end}/\Mpl = \sqrt{6} \alpha_4 \ln [1+ 1/(\sqrt{3} \alpha_4) ] $; during inflation, $\phi$ takes larger value than $ \phi_{\rm end} $.
During reheating, $\dot{\phi}$ takes the largest value at the bottom of the potential $\dot{\phi}\sim \sqrt{V(\phi_{\rm{end}})} \sim\Mpl H $.
Thus we approximately have 
\begin{equation}
    \left| \frac{4\zeta_{32}}{ \sqrt{3}} \right| \qty(1+\frac{1}{\sqrt{3} \alpha_4}) \lesssim \frac{\Mpl}{H}
~~ \text{or}~~
    \left| \frac{4\zeta_{32}}{ \sqrt{3}} \right| \qty(1+\frac{1}{\sqrt{3} \alpha_4}) \lesssim \frac{\Mpl^2}{H^2}~.
\end{equation}
For fixed value of $\zeta_{32}$, this gives the lower bound of the value of $\alpha_4$.
For fixed value of $\alpha_4$, this gives the upper bound of $\abs{\zeta_{32}}$.

Next we consider constraints from the gauge field production.
Note that Weyl transformation $g_{{\rm{E}}\mu\nu} = \Omega^2 g_{{\rm{J}}\mu\nu}$ brings an extra term from trace anomaly\footnote{Weyl transformation in curved spacetime also has anomalous terms including $R^2$, $R_{\mu\nu}R^{\mu\nu}$, etc. In the current model the coefficients of such terms are of order one and thus they are negligible compared to other terms in the action.
}
\begin{equation}
    \int \sqrt{-g_\text{E}} \dd^4 x \qty(- \frac{1}{2} \ln \Omega^2 \frac{1}{24 \pi^2 } F^{\mu \nu} F_{\mu \nu})~.
\end{equation}
We thus have
\begin{align}
S_{\rm{E}} \supset \int \sqrt{-g_\text{E}} \dd^4 x 
\left[ -\frac{1}{4} \qty(1+ \frac{\sqrt{6}\phi}{72 \pi^2 \alpha_4 \Mpl}) F^{\mu \nu} F_{\mu \nu} -\frac{1}{4} 
 \zeta_{31} \alpha_4 \left(
    e^{\frac{\phi}{\sqrt{6} \alpha_4 \Mpl}} -1 \right)
    E^{\mu \nu \alpha \beta} F_{\mu \nu} F_{\alpha \beta}\right]~.
\end{align}
To gain an analytic insight of gauge field production, we restrain our analysis to the parameter range satisfying Eqs.~\eqref{eq-epsilonB} and \eqref{eq-epsilonC} 
(see Appendix~\ref{app:gauge} for a brief summary of gauge field production during inflation). 
Specifically, 
\begin{equation}
    \epsilon_B = \frac{\Mpl^2}{2} \qty( \frac{\sqrt{6}}{72 \pi^2 \alpha_4 \Mpl + \sqrt{6} \phi})^2 \ll 1, \quad 
    \epsilon_C = \frac{1}{12 \alpha_4^2} \ll 1, \quad
    |\eta_C| = \frac{1}{6 \alpha_4^2} \ll 1
\end{equation}
is required to have an analytical result of gauge field production. 
This puts a lower bound on $\alpha_4$, \textit{i.e.}, lower bound on the prediction of tensor-to-scalar ratio at the CMB scale. 
However, note that these conditions do not restrict the model, but only the validity of the analysis method.
Non-Gaussianity imposes a constraint on the $\xi$ parameter Eq.~\eqref{eq-xiparameter} at CMB scale\cite{Barnaby:2010vf, Barnaby:2011vw}, and in this case
\begin{align}
    \xi =  \frac{12\pi^2 \zeta_{31}  e^{\frac{\phi}{\sqrt{6} \alpha_4 \Mpl}} }{\left( e^{\frac{\phi}{\sqrt{6} \alpha_4 \Mpl}} -1 \right) \left( 72\pi^2 \alpha_4 +\sqrt{6} \frac{\phi}{\Mpl} \right) }  ~,
\end{align}
$\abs{\xi}$ grows toward the end of inflation.
It may evade the non-Gaussianity bound at CMB scale whereas allows fermion production at later stage \cite{Domcke:2018eki}.

\section{Conclusions and discussion}
\label{sec:conclusion}

In this work, we have studied the couplings between matter currents and torsion in EC gravity, which is important for future discussion of particle production in EC framework, such as reheating, baryogenesis, etc. 
As concrete examples, we focus on models constructed geometrically by Ricci scalar and components of torsion within EC framework, with operators up to dimension four. 
These models generally feature an induced (pseudo-)scalar degree of freedom, the (pseudo-)scalaron, as shown in previous studies. 
In particular, such construction results in a canonicalized scalaron when the dimension-four part of the action can be written as a complete square. 
We have considered both gauge-invariant and gauge-dependent currents coupled with torsion in these models, respectively, and investigated the form of couplings with the scalaron in the equivalent metric theories.
In particular, the gauge-invariant currents can be originated from the vanilla torsion coupling with the chiral fermions.
Since matters in the standard model of particle physics are chiral fermions, the scalaron-current coupling can be regarded as a general outcome of the EC gravity framework.

In order to classify the possible realizations of the scalaron potential and scalaron-current couplings, we have introduced a general framework under the assumption that the scalaron kinetic term can be canonicalized, \textit{i.e.}, the scalaron does not exhibit the $P(\phi,X)$ type theories.
Allowing general parameters, we have studied the asymptotic behaviors of the resulting scalaron-current couplings and the scalaron potential in both large- and small-field regimes.
For the sake of completeness, we also discuss the possibilities of realizing flat potential out of these regimes, which is originated from the pole structure of the scalaron kinetic term.
Two types of models based on the large scalaron asymptotic behavior and the pole structure of the scalaron kinetic term are presented to illustrate our results.
As an example of the gauge-invariant current, we consider the chiral current coupling with the scalaron and discuss its implications for reheating.
We then discuss the Chern--Simons coupling as an example of the gauge-dependent current coupling to the scalaron, where the gauge field can be efficiently produced by the instabilities.

As a side remark, we have discussed the connection between our results and the QCD $ \theta $ term. 
In particular, we have clearly pointed out that, in the case of gauge-invariant currents, the coupling between scalaron and the currents cannot contribute to the $ \theta $ term in the QCD sector. 
On the other hand, the coupling with gauge-dependent currents could be related to the $ \theta $ term because the scalaron can couple with the Chern--Simons term. 
However, to really make the scalaron as the axion to solve the strong CP problem, the scalaron should not couple with other fields or dimensionful parameters which generally affect the effective potential of the scalaron.
The resulting potential is not guaranteed to have a desired minimum to solve the strong CP problem, which actually introduces another fine-tuning problem if we force it to be the QCD axion. 
The construction of a viable model where the scalaron can play a role as QCD axion is beyond the scope of this paper and we leave it for future work.

\vspace{0.5cm}
\noindent\textbf{Note added:} 

\noindent During the final stage of writing of this paper, the paper~\cite{Karananas:2025ews} appeared on arXiv discussing on a similar topic, \textit{i.e.}, the current couplings with the pseudo scalar field(s) induced from torsion in EC formalism.
Motivated by the previous works~\cite{Mielke:2006zp,Mercuri:2009zi,Mercuri:2009zt,Lattanzi:2009mg,Castillo-Felisola:2015ema,Karananas:2018nrj,Karananas:2024xja}, which claimed that the scalar degree of freedom originated from the torsion in EC gravity can be identified as QCD axion, the authors of \cite{Karananas:2025ews} critically discussed possible realizations.
They found that the vanilla model where the current coupling is solely originated from the torsion does not work, essentially reaching the same conclusion as ours.
They also discussed the possible addition of non-trivial interactions between QCD and the pseudo scalar field in the Weyl invariant setup.
Instead, our paper provides concrete ways for the torsion scalar degrees of freedom to couple with \emph{both} gauge-invariant and gauge-dependent currents in the EC models.
We have demonstrated explicitly that there is a gauge-invariant way to include the latter, which results in the desired form $ \propto \phi F \tilde F$ in the equivalent metric theory.
As a result, we have shown that the scalaron can be related to the QCD $\theta$ terms, yet the scalaron coupling to the dimensionful parameters is problematic for the scalaron to be a viable QCD axion.
The no-scale type generalization~\cite{Shaposhnikov:2008xi,Hong:2025tyi,Karananas:2025ews} is needed to avoid this issue, which is beyond the scope of this paper.

\section*{Acknowledgement}
M.\,He was supported by IBS under the project code, IBS-R018-D1.
M.\,Hong was supported by Grant-in-Aid for JSPS Fellows 23KJ0697.
K.\,M.\, was supported by JSPS KAKENHI Grant No.\ JP22K14044.
M. Hong gratefully acknowledges the hospitality of DESY theory group during the drafting of this paper.

\appendix

\section{Einstein--Cartan gravity}
\label{app-EC}
This appendix supplements details for Sec.~\ref{sec:pre} and settles certain conventions. 
Consider vierbein fields $e^A_\mu(x)$ and spin connection fields $\bar{\omega}^{AB}_\mu(x)$ as illustrated at the beginning of Sec.~\ref{sec:pre}.
The covariant derivative of a vector in the general coordinate and that in the local normal coordinate, respectively, is as follows:
\begin{equation}
    \bar{\nabla}_\mu V^\alpha = \partial_\mu V^\alpha + \bar{\Gamma}^\alpha_{\mu\nu}V^\nu, \quad
    \bar{\nabla}_\mu V^A = \partial_\mu V^A + \bar{\omega}^A_{\mu B}V^B~.
\end{equation}
For the compatibility of these two expressions, noting $V^A = e^A_\mu V^\mu$, one has
\begin{equation}
    \bar{\Gamma}^\rho_{\mu\nu} = e^\rho_A (\partial_\mu e^A_\nu + \bar{\omega}^A_{\mu B} e^B_\nu)~,
\end{equation}
\textit{i.e.}, $\bar{\nabla}_\mu e^A_\nu = 0$.

Define curvature $\bar{R}^A{}_{B \mu \nu}$ and torsion $T^A{}_{\mu\nu}$ as follows:
\begin{equation}
    [\bar{\nabla}_\mu,\bar{\nabla}_\nu] V^A = \bar{R}^A{}_{B \mu \nu} V^B - T^B{}_{\mu\nu} \bar{\nabla}_B V^A~,
\end{equation}
one thus have
\begin{equation}
    \bar{R}^{AB}{}_{\mu\nu} = \partial_\mu \bar{\omega}^{AB}_{\nu} -\partial_\nu \bar{\omega}^{AB}_{\mu} + \bar{\omega}^{A}_{\mu C} \bar{\omega}^{CB}_{\nu} - \bar{\omega}^{A}_{\nu C} \bar{\omega}^{CB}_{\mu}~,
\end{equation}
\begin{equation}
    T^A{}_{\mu\nu} = \partial_\mu e^A_{\nu} - \partial_\nu e^A_{\mu} + \bar{\omega}^A_{\mu B}e^B_\nu - \bar{\omega}^A_{\nu B}e^B_\mu~.
\end{equation}
One can check that $\bar{R}^{\rho}{}_{\sigma\mu\nu}=e^{\rho}_A e^B_{\sigma}\bar{R}^A{}_{B \mu \nu}$ and $T^\lambda{}_{\mu\nu} = e^\lambda_AT^A{}_{\mu\nu}$ for
\begin{equation}
    \bar{R}^{\rho}{}_{\sigma\mu\nu} = \partial_\mu \bar{\Gamma}^\rho_{\nu\sigma} - \partial_\nu \bar{\Gamma}^\rho_{\mu\sigma} + \bar{\Gamma}^\rho_{\mu \lambda} \bar{\Gamma}^\lambda_{\nu \sigma}
    - \bar{\Gamma}^\rho_{\nu \lambda} \bar{\Gamma}^\lambda_{\mu \sigma}
\end{equation}
and
\begin{equation}
    T^\lambda{}_{\mu \nu} = \bar{\Gamma}^\lambda{}_{\mu \nu} - \bar{\Gamma}^\lambda{}_{\nu \mu}~.
\end{equation}
Note that $g_{\alpha\beta} = e^A_\alpha e^B_\beta \eta_{AB}$, thus the metricity condition $\bar{\nabla}_\mu g_{\alpha\beta} = 0$ implies that $ \bar \omega^{AB}_\mu = - \bar \omega^{BA}_\mu$.
If also imposing the torsionless condition $T^A{}_{\mu\nu} = 0$, one has
\begin{equation}
    \omega^{AB}_{\mu} = \frac{1}{2} \qty[
    e^{\nu A}(\partial_\mu e^{B}_\nu - \partial_\nu e^{B}_\mu)
    - e^{\nu B}(\partial_\mu e^{A}_\nu - \partial_\nu e^{A}_\mu)
    - e_{\mu C} e^{\nu A} e^{\lambda B} (\partial_\nu e^{C}_\lambda - \partial_\lambda e^{C}_\nu)
    ]~
\end{equation}
in the mertic formalism.
To derive this one needs to notice a property for a general quantity with 3 indices:
\begin{equation}
    4 \omega^{ABC} = \omega^{[AB]C} - \omega^{[BC]A} + \omega^{[CA]B} + \omega^{A(BC)} + \omega^{B(CA)} - \omega^{C(BA)}~.
\end{equation}
where the antisymmetrization and symmetrization of the indices are defined as $\omega^{[AB]C} \equiv (1/2) (\omega^{ABC} - \omega^{BAC})$ and $\omega^{(AB)C} \equiv (1/2) (\omega^{ABC} + \omega^{BAC})$.
It is possible to check that
\begin{align}
\begin{split}
    \Gamma^\rho_{\mu\nu} 
    & = e^\rho_A (\partial_\mu e^A_\nu + \omega^A_{\mu B} e^B_\nu) \\
    & = \frac{1}{2} g^{\rho \sigma} (\partial_\mu g_{\sigma \nu} + \partial_\nu g_{\sigma \mu} - \partial_\sigma g_{\mu \nu})~.
\end{split}
\end{align}

As illustrated in Sec.~\ref{sec:pre}, in the EC gravity, metricity condition is required while torsionless condition is not.
The contorsion expresses deviation of the connections in the EC gravity from those in the metric formalism:
\begin{equation}
    \bar{\omega}^{AB}_\mu = \omega^{AB}_\mu + K^{AB}_\mu, \quad
    \bar{\Gamma}^{\rho}_{\mu\nu} = \Gamma^{\rho}_{\mu\nu} + K^{\rho}{}_{\mu\nu}~.
\end{equation}
From the metricity condition $\bar{\nabla}_\mu g_{\alpha\beta} = 0$, one has
\begin{equation}
    K^\lambda{}_{\mu\nu} = \frac{1}{2} (T^\lambda{}_{\mu\nu} + T_\nu{}^\lambda{}_\mu + T_\mu{}^\lambda{}_\nu)~.
\end{equation}
Note the property $K^\lambda{}_{\mu\nu} = - K_{\nu\mu}{}^\lambda$, which can be useful when deriving (\ref{eq-Rbar}).
One may check $K^{AB}_\mu = e^{\alpha A} e^{\beta B} K_{\alpha \mu \beta}$.
Now we decompose the torsion tensor into vector, axial vector and tensor components:
\begin{align}
    \begin{split}
        & \text{(vector)}\quad
        T_\mu = T^\nu{}_{\mu \nu}, \\
        & \text{(axial vector)}\quad
        S^\nu = E^{\rho \sigma \mu \nu} T_{\rho \sigma \mu}, \\
        & \text{(tensor)}\quad
        q_{\rho \nu \mu} = T_{\rho \nu \mu} - \frac{2}{3} T_{[\nu}g_{\mu] \rho} + \frac{1}{6}E_{\rho \nu \mu \sigma} S^\sigma~,
    \end{split}
\end{align}
where $E^{\rho \sigma \mu \nu}$ is the Levi-Civita tensor, with $E^{\rho \sigma \mu \nu} = \epsilon^{\rho \sigma \mu \nu}/\sqrt{-g}$ and $E_{\rho \sigma \mu \nu} = - \epsilon_{\rho \sigma \mu \nu} \sqrt{-g}$, and $\epsilon^{0123} = \epsilon_{0123} = 1$.
Note the properties $E^{\rho \sigma \mu \alpha} E_{\rho \sigma \mu \beta} = -6 \delta^\alpha_\beta$ and $q_{\rho \nu \mu} + q_{\nu \mu \rho} + q_{\mu \rho \nu} = 0$, which can be useful when deriving (\ref{eq-Rbar}).
The contorsion tensor can thus be expressed as
\begin{equation}
K_{\alpha \mu \beta} = \frac{2}{3} T_{[\alpha} g_{\beta]\mu} - \frac{1}{12} E_{\alpha \mu \beta \nu} S^{\nu} - q_{\mu \beta \alpha}~.
\end{equation}

In the final part of this appendix we derive (\ref{eq-naturalcoupling}).
We start from
\begin{align}
\begin{split}
    S_f & = \int \sqrt{-g} \dd^4 x\,
    \qty( \frac{i}{2} \overline{\psi} \gamma^\mu \bar{D}_\mu \psi
    -  \frac{i}{2} \overline{ \bar{D}_\mu\psi} \gamma^\mu  \psi) \\
        & = \int \sqrt{-g} \dd^4 x\,
        \qty( \frac{i}{2} \overline{\psi} \gamma^\mu D_\mu \psi-  \frac{i}{2} \overline{ D_\mu\psi} \gamma^\mu  \psi
    - \frac{i}{2} \overline{\psi} \left\{
    \gamma^\mu, \frac{1}{8} K^{AB}_\mu [\gamma_A,\gamma_B] \right\} \psi)~.
\end{split}
\end{align}
Now note the property
\begin{equation}
    \gamma_A \gamma_B \gamma_C = - \eta_{AB} \gamma_C - \eta_{BC} \gamma_A + \eta_{AC} \gamma_B + i \epsilon_{ABCD}\gamma^D \gamma_5~,
\end{equation}
where $\epsilon_{0123} = 1$ and $\gamma_5 \equiv i \gamma^0 \gamma^1 \gamma^2 \gamma^3 = -i \gamma_0 \gamma_1 \gamma_2 \gamma_3$.
Using this one can show that $(i/2)\{\gamma_A,[\gamma_B,\gamma_C]\}=-2\epsilon_{ABCD} \gamma^D \gamma_5$.
Also, $\gamma^\mu K^{AB}_\mu = \gamma_C K^{ACB} = \gamma_C[(2/3)T^{[A}\eta^{B]C} - (1/12) \epsilon^{ACBD}S_D - q^{CBA}]$.
Thus,
\begin{equation}
     - \frac{i}{2} \overline{\psi} \left\{
    \gamma^\mu, \frac{1}{8} K^{AB}_\mu [\gamma_A,\gamma_B] \right\} \psi 
    = - \frac{1}{8} \overline{\psi} \qty(-2 \epsilon_{CABD}\gamma^D\gamma_5)
    \qty(-\frac{1}{12} \epsilon^{ACBE}S_E) \psi
    = \frac{1}{8} \overline{\psi} \gamma^\mu \gamma_5 \psi S_\mu~,
\end{equation}
noting $\epsilon_{ABCD} \epsilon^{ABCE} = 6 \delta^E_D$.

\section{Chiral rotation}
\label{app-rotation}
In this Appendix we summarize the results of the chiral rotation of a Dirac fermion.
Let us start from the kinetic term of a Dirac fermion minimally coupled to gravity and charged by a U(1) gauge field.
We consider Einstein gravity here.
The action is
\begin{equation}
    S = \int \sqrt{-g} \dd^4 x\,
    \qty[
    i \bar{\psi} \gamma^\mu 
    \qty(\partial_\mu - \frac{1}{8} \omega^{AB}_{\mu} [\gamma_A,\,\gamma_B] - i g A_\mu) \psi
    ]~,
\end{equation}
where $A_\mu$ is a U(1) gauge field.

Now we perform chiral rotation:
\begin{equation}
    \psi \to 
    e^{- i \alpha(x) \gamma_5} \psi, \quad
    \bar{\psi} \to \bar{\psi} 
    e^{- i \alpha(x) \gamma_5}~,
\end{equation}
and thus
\begin{equation}
    i \bar{\psi} \gamma^\mu \partial_\mu \psi \to
    i \bar{\psi} \gamma^\mu \partial_\mu \psi 
    + j^{\mu}_5 \partial_\mu \alpha~,
\end{equation}
with $j^{\mu}_5 \equiv \bar{\psi} \gamma^\mu \gamma_5 \psi$.
If there is a mass term, it transforms as
\begin{equation}
    -m \bar{\psi} \psi \to -m\bar{\psi} e^{-2i\alpha(x)\gamma_5}\psi~.
\end{equation}
One must also take into account the Jacobian of the measure of path integral, which is related to the chiral anomaly from the gauge field and gravity \cite{Fujikawa:1980eg}.
Here we only give the final results.
The action after chiral rotation is
\begin{align}
    \begin{split}
    S = \int \sqrt{-g} \dd^4 x\,
    & \left[
    i \bar{\psi} \gamma^\mu 
    \qty(\partial_\mu - \frac{1}{8} \omega^{AB}_{\mu} [\gamma_A,\,\gamma_B] - i g A_\mu) \psi
    \right. \\
    & \left.
    - \alpha(x) \nabla_\mu j^{\mu}_5
    - \alpha(x) \frac{g^2}{16 \pi^2} E^{\mu \nu \alpha \beta} F_{\mu \nu} F_{\alpha \beta}
    + \alpha(x) \frac{1}{384 \pi^2} E^{\mu \nu \alpha \beta} R_{\rho \sigma \mu \nu} R^{\rho \sigma}{}_{\mu \nu}
    \right]~.
    \end{split}
    \label{eq-anomaly}
\end{align}
The result is easily generalized to non-Abelian gauge fields.

\section{Chiral gravitational waves}
\label{app:chiralGW}
In this appendix we briefly summarize the constraint from ghost instabilities of chiral gravitational waves from the $f(\phi)R\tilde{R}$ coupling (see \textit{e.g.} \cite{Satoh:2007gn}).
We use the following metric
\begin{equation}
    \dd s^2 = a^2(\eta) \qty[-\dd \eta^2 + (\delta_{ij} + h_{ij}(\eta,\,\vec{x})) \dd x^i \dd x^j]
\end{equation}
to calculate the second order perturbation of the action below:
\begin{equation}
    S = \int \sqrt{-g} \dd^4 x\, 
    \qty[
    \frac{\Mpl^2}{2} R
    + \frac{1}{2} f(\phi)
    E^{\mu \nu \alpha \beta}
    R_{\mu \nu \rho \sigma}
    R_{\alpha \beta}{}^{\rho \sigma}
    ]~,
\end{equation}
and obtain
\begin{equation}
    S^{(2)} = \int \dd \eta \dd^3 x\, 
    \left\{
    \frac{\Mpl^2}{8} a^2 \qty[(h'_{ij})^2 - ( h_{ij,l})^2 ]
    + \frac{\partial f}{\partial \phi }\phi' \epsilon^{ijk}
    (h'_{in} h'_{kn,j} - h_{im,n}h_{mk,jn})
    \right\}~,
\end{equation}
with 
the convention $\epsilon^{0123} = \epsilon^{123} = 1$.

Consider $(\hat{\vec{u}},\,\hat{\vec{v}},\,\hat{\vec{k}})$ with $\hat{\vec{u}} \equiv \vec{u}/|\vec{u}|,~\hat{\vec{v}} \equiv \vec{v}/|\vec{v}|$ and $\hat{\vec{k}} \equiv \vec{k}/|\vec{k}|$ which are orthogonal to each other and forms a right-handed system, and define polarization tensors
\begin{equation}
    p^+_{ij}(\hat{\vec{k}}) \equiv \hat{u}_i \hat{u}_j - \hat{v}_i \hat{v}_j, \quad
    p^{\times}_{ij}(\hat{\vec{k}}) \equiv \hat{u}_i \hat{v}_j + \hat{v}_i \hat{u}_j~.
\end{equation}
One may check the following properties:
\begin{equation}
    p^{+,\times}_{ij} = p^{+,\times}_{ji}, \quad
    k_i p^{+,\times}_{ij} = 0, \quad
    p^{+,\times}_{ii} = 0, \quad
    p^a_{ij} p^b_{ij} = 2 \delta^{ab}, \quad
    p^{+,\times}_{ij} (\hat{\vec{k}}) = p^{+,\times}_{ij} (-\hat{\vec{k}})~,
\end{equation}
with $a,~b = +$ or $\times$.
It is convenient to define circular polarization tensors:
\begin{equation}
    p^{\rm{R}}_{ij} \equiv \frac{1}{\sqrt{2}} (p^+_{ij} + i p^{\times}_{ij}), \quad
    p^{\rm{L}}_{ij} \equiv \frac{1}{\sqrt{2}} (p^+_{ij} - i p^{\times}_{ij})~.
\end{equation}
The following properties hold:
\begin{equation}
    p^{*A}_{ij} p^{B}_{ij} = 2 \delta^{AB}, \quad
    \hat{k}_l \epsilon^{mlj} p^A_{ij} (\hat{\vec{k}}) = -i \lambda^A_{\hat{\vec{k}}}p_{mi}^A(\hat{\vec{k}}), \quad
    p^{\rm{L},\rm{R}}_{ij} (- \hat{\vec{k}}) = p^{\rm{L},\rm{R}}_{ij} (\hat{\vec{k}})~,
\end{equation}
with $A,~B = $L or R and $\lambda^{\rm{R}}_{\hat{\vec{k}}} = +1,~\lambda^{\rm{L}}_{\hat{\vec{k}}} = -1$.

Decomposing the tensor modes using circular polarization tensors:
\begin{equation}
    h_{ij} (\eta, \vec{x}) = \sum_{A=\rm{R},\rm{L}}
    \int \frac{\dd^3 k}{(2 \pi)^{\frac{3}{2}}}
    p^A_{ij} (\hat{\vec{k}})
    h^A_{\vec{k}} (\eta) e^{i \vec{k} \cdot \vec{x}}~,
\end{equation}
one obtains the second-order action as
\begin{equation}
    S^{(2)} = \frac{\Mpl^2}{4} \int \dd \eta \dd^3 k\,
    \sum_{A=\rm{R},\rm{L}}
    \qty(a^2 + \frac{8 k }{\Mpl^2} \frac{\partial f}{\partial \phi} \phi' \lambda^A_{\hat{\vec{k}}})
    \qty(|h^{A \prime}_{\vec{k}}(\eta)|^2 - k^2 |h^{A}_{\vec{k}}(\eta)|^2)~.
\end{equation}
Note that the reality of $h_{ij} (\eta, \vec{x})$ implies $h^{\rm{L}}_{\vec{k}}(\eta) = \qty[h^{\rm{R}}_{-\vec{k}}(\eta)]^*$.
Absence of ghost instability thus requires
\begin{equation}
    a^2 \pm \frac{8 f' k}{\Mpl^2} > 0~.
    \label{eq-ghost}
\end{equation}

\section{Gauge field production}
\label{app:gauge}
In this Appendix, we summarize the calculation of the $\xi$ parameter of gauge field production during inflation with slowly rolling inflaton $ \varphi $  \cite{Barnaby:2011qe}.
The equation of motion of the gauge field in Coulomb gauge from variation of
\begin{equation}
    S \supset \int \sqrt{-g} \dd x^4
    \qty[
    -\frac{B(\varphi(\eta))}{4} F_{\mu\nu} F^{\mu\nu}
    - \frac{C(\varphi(\eta))}{4} F_{\mu\nu} \tilde{F}^{\mu\nu}
    ]
\end{equation}
reads
\begin{equation}
    A''_i + \frac{\partial B}{\partial \varphi} \frac{\varphi'}{B} A_i' - \frac{\partial C}{\partial \varphi} \frac{\varphi'}{B} \epsilon^{ijk}\partial_j A_k - \vec{\nabla}^2 A_i = 0~,
\end{equation}
where prime denotes derivative with $\eta$ and $ \vec{\nabla} $ is the spatial gradient operator. 

The gauge field is canonically rescaled as
\begin{equation}
    \tilde{A}_i(\eta,\vec{x}) \equiv \sqrt{B(\varphi(\eta))} A_i(\eta,\vec{x})~,
\end{equation}
which can be expanded as
\begin{equation}
    \tilde{A}_i(\eta,\vec{x}) = \sum_{\lambda = \pm} \int \frac{\dd k^3}{(2 \pi)^\frac{3}{2}} \qty[ \epsilon_i^\lambda(\vec{k}) a_\lambda(\vec{k}) \tilde{A}_\lambda(\eta,\vec{k}){\rm{e}}^{i\vec{k}\cdot\vec{x}} + \rm{h.c.}]~,
\end{equation}
where $\vec{\epsilon}^\lambda$ are circular polarization vectors.
They are defined as
\begin{equation}
    \epsilon_i^+(\vec{u})= \frac{1}{\sqrt{2}}(\hat{v}_i + i \hat{k}_i), \quad
    \epsilon_i^-(\vec{u})= \frac{1}{\sqrt{2}}(\hat{v}_i - i \hat{k}_i),~
\end{equation}
and have the following properties
\begin{equation}
    \vec{u}\cdot\vec{\epsilon}^\lambda(\vec{u}) = 0,\quad
    \epsilon^{ijk} u_j \epsilon^\pm_k(\vec{u}) = \mp i |u| \epsilon^\pm_i(\vec{u})~.
\end{equation}

The equation of motion can then be expressed using $\tilde{A}_\lambda$ as
\begin{equation}
    \qty(\frac{\partial^2}{\partial \eta^2} + k^2 + M_A^2(\eta) \pm \frac{2k\xi}{\eta})\tilde{A}_\pm(\eta,\vec{k}) = 0~,
\end{equation}
where
\begin{equation}
    \xi \equiv \sqrt{\frac{\epsilon}{2}} \Mpl \frac{\partial C}{\partial \varphi} \frac{1}{B} {\rm sign} \left( \dot{\varphi} \right)
    \label{eq-xiparameter}
\end{equation}
and 
\begin{equation}
M_A^2(\eta) = \frac{1}{4} \qty(\frac{B'}{B})^2 - \frac{1}{2} \frac{B''}{B}~.
\end{equation}
And we have used $ aH \simeq -1/\eta $ and defined the slow-roll parameter $\epsilon$ as 
\begin{equation}
    \epsilon \equiv \frac{\dot{\varphi}^2}{2 H^2 \Mpl^2}~.
\end{equation}
Note that if
\begin{equation}
    \epsilon_B \equiv \frac{\Mpl^2}{2}\qty(\frac{\partial B}{\partial \varphi} \frac{1}{B})^2 \ll 1,\quad
    |\eta_B| \equiv \left| \Mpl^2 \frac{\partial^2 B}{\partial \varphi^2} \frac{1}{B} \right| \ll 1
    \label{eq-epsilonB} ~,
\end{equation}
and
\begin{equation}
    \epsilon_C \equiv \frac{\Mpl^2}{2}\qty(\frac{\partial^2 C}{\partial \varphi^2} / \frac{\partial C}{\partial \varphi})^2 \ll 1,\quad
    |\eta_C| \equiv \left| \Mpl^2 \frac{\partial^3 C}{\partial \varphi^3} / \frac{\partial C}{\partial \varphi} \right| \ll 1~,
    \label{eq-epsilonC} 
\end{equation}
the effect of $M_A^2$ is negligible, and 
$\xi$ varies slowly and thus it can be treated as a constant \cite{Barnaby:2011qe}.


\bibliographystyle{utphys}
\bibliography{ref}

\end{document}